\documentclass[10pt]{article}

\usepackage[utf8]{inputenc}
\usepackage[english]{babel}
\usepackage{graphicx,amsmath,amsfonts,amssymb,dcolumn,amsthm,slashed,bbm,xspace,pgffor,tikz,mathbbol,latexsym,cite,braket}
\usepackage{microtype}
\usepackage{lmodern}
\usepackage[normalem]{ulem}
\usepackage{xcolor}
\usepackage[colorlinks=true,citecolor=blue,urlcolor=blue,linkcolor=blue]{hyperref}
\usepackage[hang]{footmisc} 
\usepackage[affil-it]{authblk}
\usepackage[compat=1.1.0]{tikz-feynhand}
\usepackage{multicol}
\usepackage{microtype,diagbox}

\numberwithin{equation}{section}

\usepackage{float}
\begin{document}

\title{\textbf{Confinement/deconfinement at low temperatures and rotation in the exact soft wall model}}

\author{Octavio C. Junqueira\thanks{\href{mailto:octavioj@pos.if.ufrj.br}{octavioj@pos.if.ufrj.br}}, ~ Roldao da Rocha\thanks{\href{mailto:roldao.rocha@ufabc.edu.br}{roldao.rocha@ufabc.edu.br}}}
\affil{ Center of Mathematics, Federal University of ABC, 09210-580\\ Santo Andr\'e, Brazil}

\date{}
\maketitle

\begin{abstract}

We study the effects of rotation on the confinement/deconfinement phase transition of strongly interacting matter, at low temperatures, in the soft wall AdS/QCD model at finite density. To achieve it, we apply the Hawking-Page approach to the exact Andreev's solution of a charged rotating black hole in five-dimensional AdS space. We observe that there is a critical angular velocity ($\omega_0$) of hadronic matter that depends on the baryon density, representing a strong constraint on the rotation in hadronic matter. We obtain the curve $\omega_0(\mu)$, which shows that the critical rotational velocity allowed for hadronic matter decreases as the chemical potential ($\mu$) increases. When $\mu$ approaches the most critical quark chemical potential, identified as the density of a phase transition at zero temperature for a non-rotating plasma, the rotational velocity allowed for the hadrons tends to zero. If $\omega \geq \omega_0 $, there is no phase transition and the QCD matter remains in the deconfined plasma phase. The QCD phase diagram is also obtained for the exact solution, and the critical temperatures are compared with the ones obtained from the Reissner-Nordstr\"om approximation. The results are interpreted as a consequence of contributions from regions relatively distant from the AdS boundary, which cause a non-negligible reduction in the deconfinement temperatures.

\end{abstract}

\section{Introduction}
\label{sec1}
The analysis of the quark-gluon plasma (QGP),  produced in ultrarelativistic heavy-ion collisions, has provided new important physical discoveries,  demanding in-depth theoretical and experimental developments, especially over the last decade. A crossover phase transition is well known to occur from hadronic matter to the QGP phase for quantum chromodynamics (QCD) at the Hagedorn temperature, which has garnered substantial support from experimental evidence from heavy-ion collision experiments  \cite{HotQCD:2018pds}. 
The QGP and the hadronic matter have been explored by the AdS/QCD correspondence, which emerges from the principles of the AdS/CFT correspondence \cite{Maldacena:1997re,Gubser:1998bc,Witten:1998qj}, providing a robust and insightful approach for exploring phenomena in QCD, at the strong-coupling regime \cite{Rougemont:2017tlu}. Several holographic QCD models adopt a bottom-up methodology, among which the soft wall AdS/QCD model is particularly important, mainly for its versatility in describing a plethora of phenomena in QCD   \cite{Karch:2006pv,Gursoy:2007cb}. 
In the soft wall AdS/QCD framework, weakly coupled gravity inhabits a five-dimensional anti-de Sitter (AdS${}_5$) bulk. This theoretical setup is dual to non-perturbative QCD, providing a quantitative description of strong interactions and various hadronic properties  \cite{Branz:2010ub,Colangelo:2008us,Brodsky:2014yha,Rougemont:2023gfz}. In this setup, a strongly coupled QCD can be formulated at the AdS${}_5$ boundary when considering low-energy conditions, whereas the additional spatial dimension across the AdS bulk represents the QCD energy scale. 
In the infrared (IR) limit, the bottom-up AdS/QCD correspondence plays a leading role in accurately describing fundamental features of QCD through appropriate expressions for the warped factor and the dilaton background. The strongly coupled dynamics leads to critical phenomena such as spontaneous chiral symmetry breaking and the confinement of quarks and gluons inside hadrons.  The soft wall AdS/QCD model effectively implements non-perturbative aspects of confinement, allowing for a deeper understanding of strong interactions \cite{Ballon-Bayona:2017sxa,Bartz:2018nzn,daRocha:2021xwq,Ballon-Bayona:2023zal,Toniato:2025gts,Aref'eva2018May}. 
Despite lacking an exact conformal symmetry underlying QCD, employing the principles of conformal field theories is still conceivable to construct the soft wall AdS/QCD model, establishing a dual theory with parameters that corroborate phenomenological data \cite{Bernardini:2018uuy,Braga:2020opg,Ferreira:2019inu,Shukla:2023pbp,Cucchieri:2007rg,Ferreira:2020iry,daRocha:2024lev}. The soft wall AdS/QCD model is explicitly designed by incorporating a dilaton field that is coupled to Einstein-Hilbert gravity. This procedure introduces a smooth cutoff in the AdS geometry, naturally yielding confinement, chiral symmetry breaking, and the hadronic Regge trajectories \cite{Csaki,Gherghetta:2009ac,Erlich:2005qh,Sakai2005Apr}. In particular, using AdS/QCD approaches to describe rotating plasmas, it was found that the critical temperature of  confinement/deconfinement transition decreases with increasing angular velocity, see \cite{Chen:2020ath,Braga:2022yfe, Zhao:2022uxc}, which is in agreement with the phenomenological Nambu-Jona-Lasinio model \cite{Wang:2018sur}, and also  with the hadron resonance gas one \cite{Fujimoto:2021xix}.

In the fluid/gravity correspondence, the Navier-Stokes equations that govern viscous relativistic hydrodynamics are dual to Einstein's field equations, which describe how gravity operates in the AdS bulk. Operators associated with  graviton fields in the AdS bulk are dual to the energy-momentum tensor regulating  the strong-coupled quantum field theory on the AdS  boundary \cite{Bemfica:2020xym,Rocha:2022ind,Kovtun:2019hdm}. In the long-wavelength IR limit, the fact that the energy-momentum tensor is conserved leads to a hydrodynamical dual description. In this way, gravity in the AdS bulk can be related to viscous relativistic hydrodynamics on the AdS boundary \cite{Bhattacharyya:2007vjd, Haack:2008cp, Policastro:2002se}.
This apparatus can be used within AdS/QCD. In ultrarelativistic heavy-ion collisions, for nuclei undergoing off-axis collisions rather than head-on ones, rotation naturally sets in the QGP, providing a substantial amount of angular momentum \cite{STAR:2017ckg}.  Therefore, the QGP acquires vorticity, enhancing the elliptic flow. From the point of view of hydrodynamics, the initial angular momentum encodes a non-trivial
dependence of the initial (longitudinal) fluid flow velocity, implementing vorticity in the Navier-Stokes equations of motion,
increasing the QGP expansion rate, and absorbing the quenching effect of shear and bulk viscosities  \cite{Abboud:2023hos}. The QGP vorticity 
has physical signatures, as  generation of an average polarization of the emitted
hadrons and the chiral vortical effect \cite{Jiang:2016woz,Becattini:2007sr,Krein:2017usp,Kharzeev:2015znc,GoncalvesdaSilva:2017bvk}.
A deeper understanding of rotation effects in the QGP can shed new light on strongly coupled systems in QCD. For it, several relevant approaches contributed to important developments in the context of AdS/QCD \cite{Braga:2022yfe,Braga:2023qee}. 
In this work, the effects of rotation on the confinement/deconfinement phase transition of
strongly interacting matter are scrutinized at low temperatures and finite
density. Andreev’s solution
of a charged rotating black hole in five-dimensional AdS spacetime is employed \cite{Andreev:2010bv}, 
whose Hawking-Page transition,  between the thermal AdS and AdS black holes with radiation, is studied. 
It gives rise to a dependence of the critical angular velocity
of hadronic matter on both the baryon density and the chemical potential of the hot medium in QCD, representing a strong constraint on the rotation of hadronic matter. One of the main results 
reveals that when the chemical potential tends to the most critical quark chemical
potential, rotation effects vanish. For specific ranges of the angular velocity of the QGP, no
phase transition is observed, leading QCD matter to correspond to the deconfined QGP phase. The QCD phase diagram is also obtained for Andreev's exact solution, whose critical temperatures are also analyzed and compared to the ones in the Reissner-Nordstr\"om limit.

This work is organized as follows: in Section \ref{sec2},  the rotating charged AdS black hole is introduced and its Hawking temperature is shown to provide an upper bound for the black hole event horizon, yielded by the black hole charge. In Section \ref{sec3}, the regularized black hole action is implemented in the soft wall AdS/QCD model. In this framework, Andreev’s solution 
is addressed and discussed. 
The critical values of the angular momentum  of hadronic matter are presented in Section \ref{sec4} together with the critical temperatures of confinement/deconfinement
transition at different quark chemical potentials. 
Section \ref{sec5} is devoted to analyzing the QCD phase diagram with respect to the quark-chemical potential and rotation.  
The Hawking-Page curve of the system is obtained. Section \ref{sec6} displays the conclusions and final remarks.

\section{Rotating charged AdS black hole geometry}
\label{sec2}
For a holographic description of a rotating quark-gluon plasma at finite temperature with non-zero quark chemical potential, the gravitational dual is a charged black hole with non-zero angular momentum, in five-dimensional AdS spacetime, which is a 
solution of Einstein's equations with a negative cosmological constant $ \Lambda = - \frac{12}{L^2} $,  and constant curvature $R = - \frac{20}{L^2}  $, where $L$ is the curvature radius of AdS. Assuming a plasma with cylindrical symmetry, rotating with a uniform angular velocity $\omega$ around a hypercylinder with radius $l$,  it can be described by the following black hole (BH) metric in the canonical form
\cite{Zhou:2021sdy, LEMOS199546, BravoGaete:2017dso}:
\begin{eqnarray}\label{canon}
	ds^2 = N(z,q) dt^2 + \frac{L^2}{z^2}\frac{dz^2}{f(z,q)} + R(z,q)\left( d\phi + P(z,q) dt\right)^2 + \frac{L^2}{z^2} \sum_{i=1}^2 dx_i^2\;,
\end{eqnarray}
with 
\begin{eqnarray}
	N(z,q) &=& \frac{L^2}{z^2} \frac{ f(z,q) (1-\omega^2 l^2)}{1- f(z,q)\omega^2 l^2}\;, \\
	R(z,q) &=& \frac{L^2}{z^2}\left( \gamma^2 l^2 -  f(z,q) \gamma^2 \omega^2 l^4\right)\;, \\
	P(z,q) &=& \frac{\omega(1-f(z,q))}{1- f(z,q)\omega^2 l^2}\;,
\end{eqnarray}
where $\gamma = 1/\sqrt{1 - l^2 \omega^2 }$ stands for the Lorentz factor, and 
\begin{equation}\label{f(z)}
f(z,q) = 1 - \frac{z^4}{z_h^4}-q^2 z_h^2 z^4 + q^2 z^6\;,
\end{equation}
being $z_h$ the location of the black hole event horizon, such that $f(z_h,q) = 0$, whereas $q$ denotes a parameter that is proportional to the BH charge. On the other hand, the gauge dual of the hadronic phase is given by the thermal AdS spacetime, which is described by the same metric \eqref{canon}, taking the limit $f(z,q) \rightarrow 1$. 

The Hawking temperature of the rotating charged BH can be obtained from the surface gravity formula. Defining $h_{00}(z) \equiv - N(z)$  Ref. \cite{Zhou:2021sdy}, one has
\begin{eqnarray}\label{HTrot}
	T(q, \omega) &=& \left\vert \frac{\kappa_G}{2\pi} \right\vert = \frac{1}{2\pi}\left\vert \lim_{z\rightarrow z_h}- \frac{1}{2} \sqrt{\frac{g^{zz}}{-h_{00}(z)}}h_{00,z}\right \vert = \frac{1}{\pi z_h} \left(1-\frac{{q}^2 z_h^6}{2}\right) \sqrt{1-\omega^2 l^2}\;, 
    \end{eqnarray}
where $\kappa_G$ is the surface gravity, and $g^{zz}$ denotes the $z-z$ component of the inverse of the cylindrical BH metric \eqref{canon}. 
 In the gauge/gravity duality dictionary, the Hawking temperature emulates the temperature of the hydrodynamical fluid flow in thermal equilibrium, describing the QGP.  
One observes the existence of a physical condition for the positivity of the temperature, given by  
\begin{equation}\label{positivity}
  z_h \leq \left(\sqrt{2}/q\right)^{1/3}\;.
\end{equation}

For this system with compactified time coordinate, the time period associated with the BH is given by $\beta = 1/T$, being $T$ the BH temperature \eqref{HTrot}.  By requiring that the asymptotic limits of the thermal and BH AdS geometries in the rotating system are the same at $z=\epsilon$, with $ \epsilon \to 0$, one finds that the thermal AdS period is given by 
\begin{equation}\label{betaAdS}
\beta_{AdS} (q,\omega) = \beta(q, \omega) \sqrt{f(\epsilon, q)}\;,
\end{equation}
which will be used to obtain the regularized BH action density, which is used to compute the confinement/deconfinement critical temperatures, defined via holography by the Hawking-Page (HP) transition between the BH and thermal AdS geometries, recently supported by quantum information entropy approaches \cite{Braga:2016wzx,Barbosa-Cendejas:2018mng}.

\section{Rotating BH exact solution in the soft wall AdS/QCD model}
\label{sec3}
We will apply the holographic soft wall AdS/QCD model \cite{Karch:2006pv}  to compute the critical temperatures of deconfinement for a rotating QGP at finite density, as a function of its angular velocity and the quark chemical potential. In this model, one introduces an energy parameter in the AdS geometry, which is interpreted as an IR cutoff on the gauge side of the duality. In Euclidean space, the five-dimensional  gravitational action of this model is given by \cite{Herzog:2006ra,BallonBayona:2007vp}
\begin{equation}\label{action1}
	I_G = - \frac{1}{ 2 \kappa^2} \int dz\int d^4x \sqrt{g} e^{-\Phi}\left( R - \Lambda \right) \;,
\end{equation}
where $\Phi(z) = cz^2$ is the  dilaton background scalar field that breaks the conformal symmetry, responsible for introducing the IR mass scale $\sqrt{c}$ into the theory, while $\kappa$ is the gravitational coupling associated with the Newton gravitational constant. The determinant of the metric, for both AdS spacetimes, reads $g = l^2 L^{10}/z^{10}$.

Thus, using the relation between the AdS curvature and the cosmological constant $\Lambda$, one obtains the gravitational on-shell action
\begin{equation}
	I_{G_{on-shell}} = \frac{4l^2L^3}{\kappa^2} V_{3D} \int_0^{\beta_s} dt \int_0^{z_h} dz\, z^{-5}e^{-cz^2}\;\,,
\end{equation}  
 where $\beta_s$ is the period of the corresponding space. For the thermal AdS geometry, one verifies that $z_h \rightarrow \infty$, since this space has no horizon. As the integration over the spatial bulk coordinates $x$ is trivial in both spaces, it generates only a volume factor $V_{3D}$. 

For a system with quarks, we must introduce an Abelian field $V_\mu $ living in AdS space, which is associated with the black hole charge. In the holographic description, the time component of $V_\mu$ works as the source of the correlation functions of the gauge theory density operator, since $V_0$ is coupled to the density $ J^0 $. This way, $V_0$ is interpreted as the quark chemical potential ($\mu$) correlated to the quark density $\bar{\psi}_\mu \gamma^0 \psi_\mu$ in the bulk. The five-dimensional action governing the vector fields is given by \cite{Braga:2015jca}
\begin{equation}\label{actionVF}
    I_{VF} = -\frac{1}{4g_5^2}\int dz \int d^4x \sqrt{g} e^{-\Phi} F_{MN}F^{MN}\;,
\end{equation}
where $F_{MN} = \partial_{M} V_N - \partial_N V_M$. To obtain an on-shell version of $I_{VF}$ for a rotating system, without employing the Reissner-Nordstr\"om (RN)  approximation, we must work with the exact solution of the equations of motion involving the gauge field in AdS space, according to the rotating charged BH metric \eqref{canon}. 

\subsection{Andreev's exact solution for the gauge field}

Andreev's exact solution for the gauge field in the soft wall model is obtained by assuming that $V_i = V_z = 0$ in the non-rotating system \cite{Andreev:2010bv}. For a charged BH with zero angular momentum, we must take the limit $\omega l \rightarrow 0$ in the metric \eqref{canon}, which yields
\begin{equation}\label{CBH}
	ds^2 = \frac{L^2}{z^2} \left( f(z,q)dt^2 + l^2 d\phi^2 + \sum_{i=1}^2 dx_i^2 + \frac{dz^2}{f(z,q)}\right)\;. 
\end{equation}
The equation of motion for the gauge field reads
\begin{eqnarray}
    \partial_P\left(e^{-\Phi} g^{M P} g^{N Q} F_{M N} \right) = 0\;, 
\end{eqnarray}
whose exact Andreev's solution for the metric \eqref{CBH} reads 
\begin{eqnarray}
   V_0 &=& A_0(z) = i \left( \frac{C_1 e^{cz^2}}{c} + C_2 \right)\;, \label{A1} \\
   V_i &=& V_z = 0\;, \label{A2}
\end{eqnarray}
where $C_1$ and $C_2$ are constants, and the $i$-factor appears due to the Wick rotation. In the limit $z \rightarrow 0$, we must obtain the solution given by the RN approximation, which is valid for small $z$, given by
\begin{eqnarray}\label{RN}
   A_0^{RN}(z) = A_0(z \rightarrow 0) = i(\mu - \eta q z^2)\;,  
\end{eqnarray}
where $\eta$ is a free parameter. From Eqs. \eqref{A1} and \eqref{RN}, one concludes that 
\begin{eqnarray}
    \mu &=& \frac{C_1}{c} + C_2 \;, \label{muC1C2}\\
     \eta q &=& -C_1\label{qC1}\;. 
\end{eqnarray}
Thus, replacing Eqs. \eqref{muC1C2} and \eqref{qC1} into Eq. \eqref{A1}, one obtains
\begin{eqnarray}\label{A01}
    A_0(z) = i \left[ \mu -\frac{\eta q}{c} \left(e^{c z^2} - 1 \right)\right]\;. 
\end{eqnarray}

The Dirichlet boundary condition $A_0(z_h) = 0$ is required to obtain a gauge field with regular norm, see Refs. 
\cite{Horigome:2006xu, Nakamura:2006xk, Hawking:1995ap, Ballon-Bayona:2020xls}, and yields the following relation between the $q$ parameter and $\mu$:
\begin{eqnarray}\label{qmu}
    \frac{\eta q}{c} = \frac{\mu}{e^{c z_h^2}-1}\;,  
\end{eqnarray}
which correlates the BH charge $Q = \eta q$ with the quark chemical potential, which is used in the holographic description of the QCD phase diagram \cite{Lee:2009bya}. In the limit of small $z_h$, one recovers the well-known relation in the RN approximation, where $\mu = \eta q z_h^2$. Replacing Eq. \eqref{qmu} into Eq. \eqref{A01}, one gets the final expression for the time component of the gauge field for the non-rotating system, 
\begin{eqnarray}\label{A0final}
    A_0(z) = i\mu \left( \frac{e^{c z_h^2} - e^{c z^2}}{e^{c z_h^2}-1}\right)\;.
\end{eqnarray}

The rotating charge BH metric \eqref{canon} can be obtained from the coordinate transformation \cite{BravoGaete:2017dso, PhysRevD.97.024034}  
\begin{eqnarray}
	t &\mapsto& \frac{1}{\sqrt{1- \omega^2 l^2}} \left(t + l^2 \omega \phi \right)\;,\label{T1}\\
	\phi &\mapsto& \frac{1}{\sqrt{1- \omega^2 l^2}} \left(\phi +  \omega t \right)\label{T2}\;,
\end{eqnarray}  
on the non-rotating metric \eqref{CBH}, which corresponds to a Lorentz boost to an observer that is rotating around the hypercylinder with radius $l$, for which the angular coordinate is varying uniformly with time, with angular velocity $\omega$. One concludes that, for the rotating system, the exact solution is given by the gauge-transformed field 
\begin{eqnarray}
    A^\prime_0 &=& \gamma(\omega l) A_0 \;,\label{A0prime}\\
    A_\phi^\prime & = &-l^2\omega \gamma(\omega l) A_0\label{Aphiprime}\\
    A_{x_1}^\prime &=& A_{x_2}^\prime = A^\prime_z = 0\;, \label{AiAzprime}
\end{eqnarray}
with $A_0$ defined by Eq. \eqref{A0final}. 

The gauge-invariant quark chemical potential in the rotating system, associated with the BH charge via duality, can be defined by the expression  \cite{Chen:2020ath,Zhao:2022uxc} 
\begin{equation}
    \mu^\prime = A_\mu^\prime \chi^\mu\vert_{z = 0} - A_\mu^\prime \chi^\mu\vert_{z = z_h}
\end{equation}    
where the Killing vector $\chi = \partial_t + \omega \partial_\phi$ is the null generator of the horizon that is rotating with angular velocity $\omega$. Using  Eqs. \eqref{A0prime} -- \eqref{AiAzprime}, and comparing the result with the static case, one finds
\begin{eqnarray}\label{muprime}
    \mu^\prime  = \mu\sqrt{1 - \omega^2 l^2} \;,
\end{eqnarray}
which shows that the chemical potential transforms as the inverse of the Lorentz factor. This result is consistent with the transformation of the interacting term $A_M J^M$, which is invariant under the coordinate transformations \eqref{T1} and \eqref{T2}. From the gauge field in the rotating system, one can compute the on-shell version of the action $I_{VF}$ and then apply the HP  approach to compute the deconfinement temperatures for the rotating QGP in the exact soft wall model. 

\subsection{Regularized rotating charged BH action density}

From the exact Andreev's solution \eqref{A0prime} -- \eqref{AiAzprime} for the gauge field in a rotating charged BH, the only non-vanishing components of the field strength {\color{cyan}{in the action \eqref{actionVF}}} are 
\begin{eqnarray}
F_{tz} &=& - F_{zt} =  2ic\gamma \mu  \frac{z e^{cz^2}}{e^{c z_h^2}-1}\;, \label{Ftz}\\   
F_{\phi z} &=& - F_{z \phi} = - 2ic\gamma \mu  \omega l^2 \frac{z e^{cz^2}}{e^{c z_h^2}-1}\;,\label{Fphiz}
\end{eqnarray}
with the non-vanishing components of the inverse of the metric \eqref{canon} given by 
\begin{eqnarray}
    g^{tt} &=& \frac{z^2\left(1+l^2 \omega^2 f(z,\mu)\right)}{L^2 f(z, \mu)} \gamma^2\;,\label{gtt} \\
    g^{1 1} &=& g^{2 2} = 1\;,\\
    g^{\phi \phi} & = & \frac{z^2\left( l^2 \omega^2 + f(z,\mu)\right)}{L^2 l^2 f(z,\mu)} \gamma^2\;,\\
    g^{zz} & = & \frac{z^2 f(z,\mu)}{L^2}\;,\\
    g^{t\phi} &=& g^{\phi t} = -\frac{z^2\left(1+f(z,\mu) \right)}{L^2 f(z,\mu)}\omega\gamma^2\;.\label{gtphi}
\end{eqnarray}
Replacing Eqs. \eqref{Ftz}, \eqref{Fphiz}, and \eqref{gtt} -- \eqref{gtphi} into Eq. \eqref{actionVF}, one obtains the on-shell $U(1)$ gauge action
\begin{eqnarray}\label{action2}
    I_{VF_{on-shell}} =  \frac{2lLc^2 \mu^2}{g^2_5(e^{cz_h^2}-1)^2 }\gamma^4 V_{3D} \, \int_0^{\beta_s} dt  \int_\epsilon^{z_{min}}dz\, z e^{-c z^2}\left[(1+l^2\omega^2)^2 + 4l^2\omega^2f(z,\mu)\right]\;.
\end{eqnarray}
In the limit of small $z_h$, one recovers the expression obtained in the RN approximation, see \cite{Braga:2025eiz}. The total on-shell action in the exact soft wall model for a rotating QCD matter at finite density is  
\begin{equation}\label{totalI}
    I_{on-shell} = I_{G_{on-shell}} + I_{VF_{on-shell}}\;. 
\end{equation}

 The trivial factor $l V_{3D}$ appears in both actions, $I_{G_{on-shell}}$ and $I_{{VF}_{on-shell}}$. For this reason, we define the total action density $\mathcal{E} = \frac{1}{lV_{3D}} I_{on-shell}$, which is given by
\begin{equation}
	\mathcal{E}_s(\varepsilon) =  \beta_s  \int_\varepsilon^{z_{min}}dz\, \frac{e^{-c z^2}}{z^5}\left[ \frac{4L^3}{\kappa^2} + \frac{2Lc^2\mu^2}{g^2_5(e^{cz_h^2}-1)^2}\gamma^4\left((1+l^2\omega^2)^2+4l^2\omega^2f(z,\mu)\right) z^6\right]\;, 
\end{equation}
where we introduced the ultraviolet (UV) regulator $\varepsilon$ in the integration over $z$. The action densities of both spaces are infinite in the limit $\varepsilon \rightarrow 0$.  The regularized black hole action density, without UV divergencies, is defined as the difference between the action densities of each space,  \begin{equation}\label{DeltaE}
	\bigtriangleup \mathcal{E}(\varepsilon) = \lim_{\varepsilon \rightarrow 0} \left[\mathcal{E}_{BH}(\varepsilon) - \mathcal{E}_{AdS}(\varepsilon) \right]\;, 
\end{equation}
with
\begin{eqnarray}
	\mathcal{E}_{BH}(\varepsilon) &=& \beta\int_\varepsilon^{z_h}dz \frac{e^{-c z^2}}{z^5}\left[ \frac{4 L^3}{\kappa^2} + \frac{2Lc^2\mu^2}{g^2_5(e^{cz_h^2}-1)^2}\gamma^4\left((1+l^2\omega^2)^2+4l^2\omega^2f(z,\mu)\right) z^6\right]  \;, \label{DeltaEBH}\\
	\mathcal{E}_{AdS}(\varepsilon) &\!=\!&  \beta \sqrt{1 \!-\! \frac{\varepsilon^4}{z^4} \!-\! \frac{c^2\mu^2(z_h^2 \varepsilon^4 \!+\! \varepsilon^6)}{\eta^2(e^{cz_h^2}-1)^2}}\!\int_\varepsilon^{\infty}\!dz \frac{e^{-c z^2}}{z^5}\left[ \frac{4L^3}{\kappa^2} \!+\! \frac{2Lc^2\mu^2}{g^2_5(e^{cz_h^2}\!-\!1)^2}\gamma^4\!\left((1\!+\!l^2\omega^2)^2\!+\!4l^2\omega^2f(z,\mu)\right) z^6\right]\;, \nonumber\\   \label{DeltaEAdS}
\end{eqnarray}
where we used the expression of $\beta_{AdS}$ \eqref{betaAdS} with $q(\mu,z_h)$ given by Eq. \eqref{qmu}. 

By defining the dimensionless variables
\begin{eqnarray}
    \bar{z}_h &=& z_h \sqrt{c}\;,\nonumber\\
    \bar{\mu} &=& \mu/\sqrt{c}\;,\nonumber\\
    \bar{q} &=& q/c^{3/2}\;,\label{varsw}
    \end{eqnarray}
and assuming $\eta = 1$ {\color{cyan}{in Eq. (\ref{RN})}}, after replacing \eqref{DeltaEBH} and \eqref{DeltaEAdS} into Eq. \eqref{DeltaE}, one obtains the final expression for the regularized BH action density: 
\begin{eqnarray}\label{deltaEsoft}
    \bigtriangleup \bar{\mathcal{E}}(\bar{\mu},\omega , \bar{z}_h) &=& \frac{e^{-\bar{z_h}^2}\pi \bar{z}_h \gamma(wl) } {  2 z_h^4\left(1-\frac{\bar{\mu}^2 \bar{z}_h^6}{2(e^{ \bar{z_h}^2}-1)^2}\right) } \left[2(-1+\bar{z_h}^2) + e^{\bar{z}_h^2}\left(1+\frac{\bar{\mu}^2 \bar{z}_h^6}{(e^{ \bar{z}_h^2}-1)^2}\right)  + 2\bar{z}_h^4e^{\bar{z_h}^2}\text{Ei}(-\bar{z}_h^2)\right.\nonumber\\
    && -\left.\frac{\bar{\mu}^2z_h^4}{(e^{ \bar{z}_h^2}-1)^2}\bar{s}_1(\omega, \bar{z}_h) +\frac{\bar{\mu}^4z_h^8}{(e^{ \bar{z}_h^2}-1)^4}\bar{s}_2(\omega, \bar{z}_h)\right]\;,\nonumber\\ 
\end{eqnarray}
where $\text{Ei}(x) = - \int_{-x}^\infty e^{-t}/t\,dt $ is the  exponential integral, and $\bigtriangleup \bar{\mathcal{E}} = \kappa^2 \bigtriangleup \mathcal{E}/(L^3 c^{3/2})$ is the dimensionless action density, according to the definitions
\begin{eqnarray}
    \bar{s}_1(\omega, \bar{z}_h) &=& \gamma^4\left[ 3 + 3l^4\omega^4 - \frac{6l^2 \omega^2}{\bar{z}_h^4}\left( 4 + 4\bar{z}_h^2 - \bar{z}_h^4\right)\right] \;,\\
    \bar{s}_2(\omega, \bar{z}_h) &=& \frac{12 l^2 \omega^2\gamma^4}{z_h^8}\left(6 + 4\bar{z}_h^2 + \bar{z}^4_h\right)\;.
\end{eqnarray}

In the holographic HP approach, the regularized BH action density \eqref{DeltaE} defines  the critical temperatures of confinement/deconfinement transition as a function of $\mu$ and $\omega$. When $\bigtriangleup \mathcal{E}$ is positive (negative), the BH is unstable (stable), since the Gibbs free energy density ($\Phi_{Gibbs} = \frac{1}{\beta}\bigtriangleup \mathcal{E}$) of the AdS space is smaller (greater) than the black hole one. In the AdS/QCD duality, the thermal AdS space corresponds to the hadronic phase, while the BH phase describes the QGP. The phase transition occurs when 
\begin{equation}\label{HPtransition}
    \bigtriangleup \mathcal{E}(\mu,
 \omega, z_h) = 0  \quad  \text{at} \quad {z_h} = {z_h}_c(\mu, \omega)\;.
\end{equation} 
To obtain the curves described by $T_c(\mu, \omega l)$, we must compute the critical horizon ${z_h}_c(\mu, \omega)$. Since Eq. \eqref{HPtransition} does not have an analytical solution, we are forced to apply numerical methods.  

\begin{figure}[!htb]
	\centering
	\includegraphics[scale=0.54]{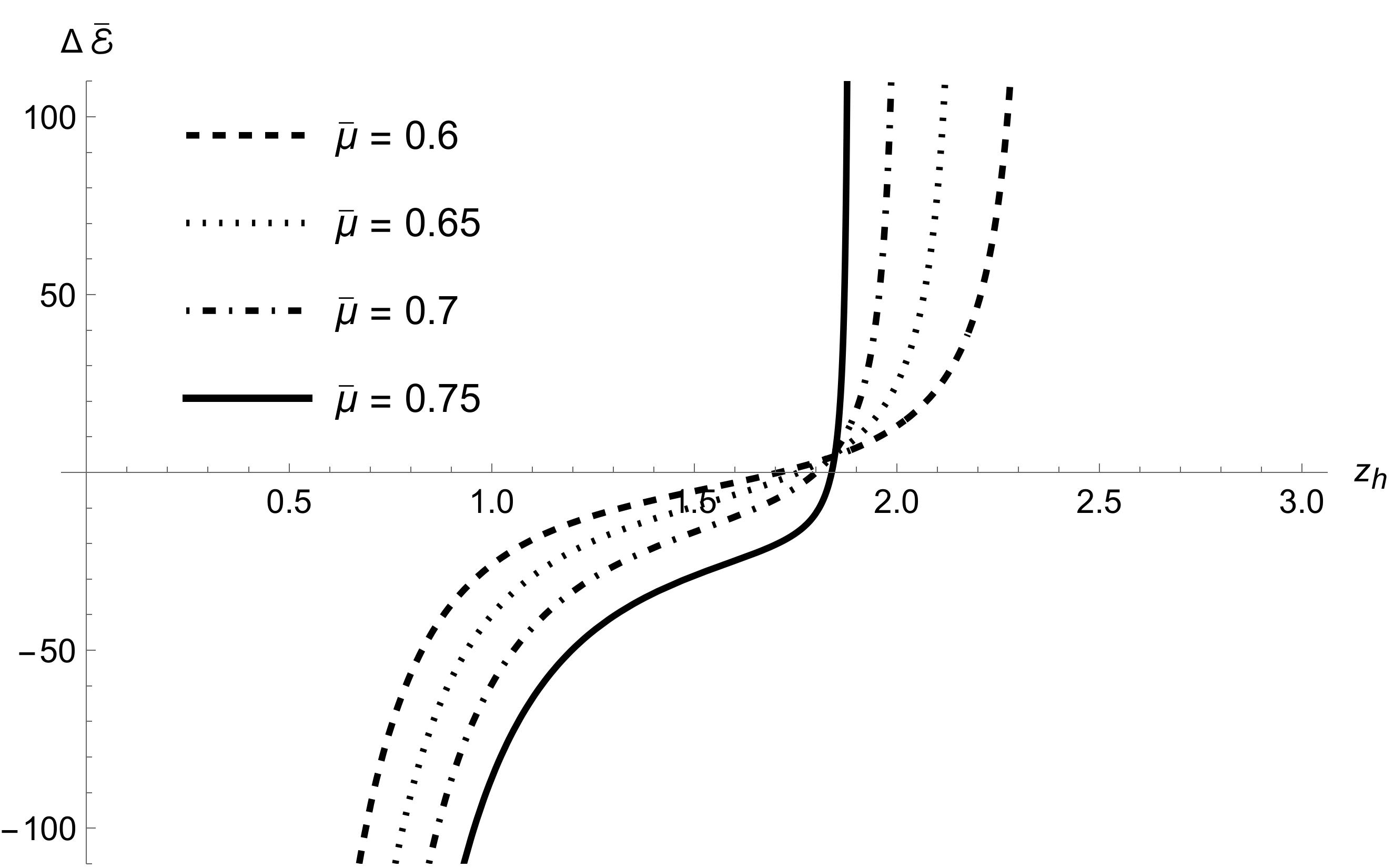}
	\caption{Action density of regularized rotating BH at finite density versus horizon position from the exact Andreev's solution of soft wall model, at a fixed plasma rotational velocity ($\omega l = 0.5$), and different quark chemical potentials.}
    \label{fig1}
\end{figure}

In Fig. \ref{fig1}, we have plotted $\bigtriangleup \bar{\mathcal{E}}$ as a function of the horizon position, at a fixed rotational velocity ($\omega l = 0.5)$, and different quark chemical potentials. The curves diverge as they approach the edge of the positivity temperature condition, $\bar{\mu}^2 \bar{z}_h^6/2(e^{\bar{z}_h} - 1)^2 \leq 1$, see Eq. \eqref{positivity}. As we can see, the critical horizons are affected by variations in the quark density. These curves are similar to the ones obtained through the RN approximation, see again \cite{Braga:2025eiz}. The critical horizon increases as the quark chemical potential increases, which corresponds to smaller critical temperatures of deconfinement, as we shall see in the next sections.      

\section{Deconfinement and the most critical rotation of hadronic matter}
\label{sec4}
For a cylindrical charged BH, where $l$ is constant, the Hawking temperature is a function of $q$ (or $\mu$), $\omega$, and $z_h$. Using Eqs. \eqref{HTrot} and \eqref{qmu}, its dimensionless version in Andreev's exact soft wall model takes the form
\begin{eqnarray}\label{barT}
\bar{T}(\bar{\mu}, \omega, \bar{z}_h) \equiv \frac{T}{\sqrt{c}}=  \frac{1}{\pi \bar{z}_h} \left(1-\frac{{\bar{\mu}}^2 z_h^6}{2(e^{\bar{z}^2_h}-1)^2}\right) \sqrt{1-\omega^2 l^2}\;. 
\end{eqnarray}

\begin{figure}[!htb]
	\centering
	\includegraphics[scale=0.52]{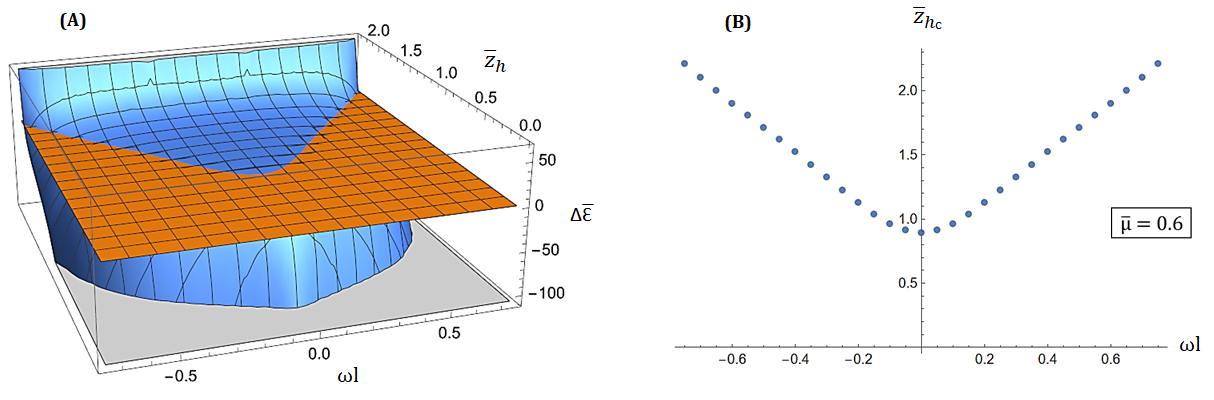}
	\caption{(A) Action density of regularized rotating BH as a function of $\bar{z}_h$ and $\omega l$, at a fixed quark chemical potential $\bar{\mu} = 0.6$, with the intersection with the plane $\Delta \bar{\mathcal{E}} = 0$. (B) Critical horizon versus plasma rotational velocity at the same fixed $\bar{\mu}$.}
    \label{fig2}
\end{figure}

We can determine the critical temperatures at a given fixed quark chemical potential, using the HP transition equation \eqref{HPtransition} together with \eqref{deltaEsoft}. In Fig. \ref{fig2}(A), we have plot $\bigtriangleup \bar{\mathcal{E}}$ as a function of $\omega$ and $z_h$ at $\mu = 0.6$. This surface represents the charged BH free Gibbs energy up to a factor $1/\beta$. Its intersection with the plane $\Delta \bar{\mathcal{E}} = 0 $, see again Fig. \ref{fig2}(A), represents the critical horizon $\bar{z}_{hc}$ as a function of $\omega$. In Fig. \ref{fig2}(B), we plot the numerical values of $z_{h c}$ as a function of $\omega l$, at $\bar{\mu} = 0.6$, see Table \ref{table1} in the Appendix, that corresponds to the curve $z_{hc}(\omega l)$ generated by the intersection in Fig. \ref{fig2}(A). For all quark chemical potentials analyzed, the behavior of $z_{h c}$ is similar, see again Table \ref{table1}.   

From the values of $\bar{z}_{hc}$ at different quark chemical potentials,  we can obtain the curves $\bar{T}_c(\omega l)$ for each fixed $\bar{\mu}$, using Eq. \eqref{barT}. The values of these critical temperatures are displayed in Table \ref{table3} in the Appendix and correspond to the points in Fig. \ref{fig3}. These lines display HP transitions for each system at $\bar{\mu}$. Below the curve, the system is in the hadronic confined phase, and above it, in the QGP one. The critical temperatures decrease as the chemical potential increases. This behavior can be obtained by applying the RN approximation, see \cite{Braga:2025eiz}. The analysis of the transitions at low temperatures, however, can only be done with high precision using the exact solution, since the values of $\bar{z}_{hc}$ escape from the RN approximation that requires $\bar{z}_h \approx 1$. In this figure, one observes that there is a maximum value of the plasma rotational velocity ($\omega_{0}$), for each curve at a given $\bar{\mu}$, above which no more phase transitions occur. For $\omega  > \omega_{0} $, the QCD matter is always in the QGP phase. The same is observed in the RN approximation, see Tables \ref{table3} and \ref{table4}, corresponding to Fig. \ref{fig4} (left). However, the critical temperatures in the exact case are lower than those generated by the RN approximation, see Fig. \ref{fig4} (right), showing that contributions from regions relatively distant from the AdS boundary induce a considerable reduction in the temperatures at different rotations.

\begin{figure}[!htb]
	\centering
	\includegraphics[scale=0.6]{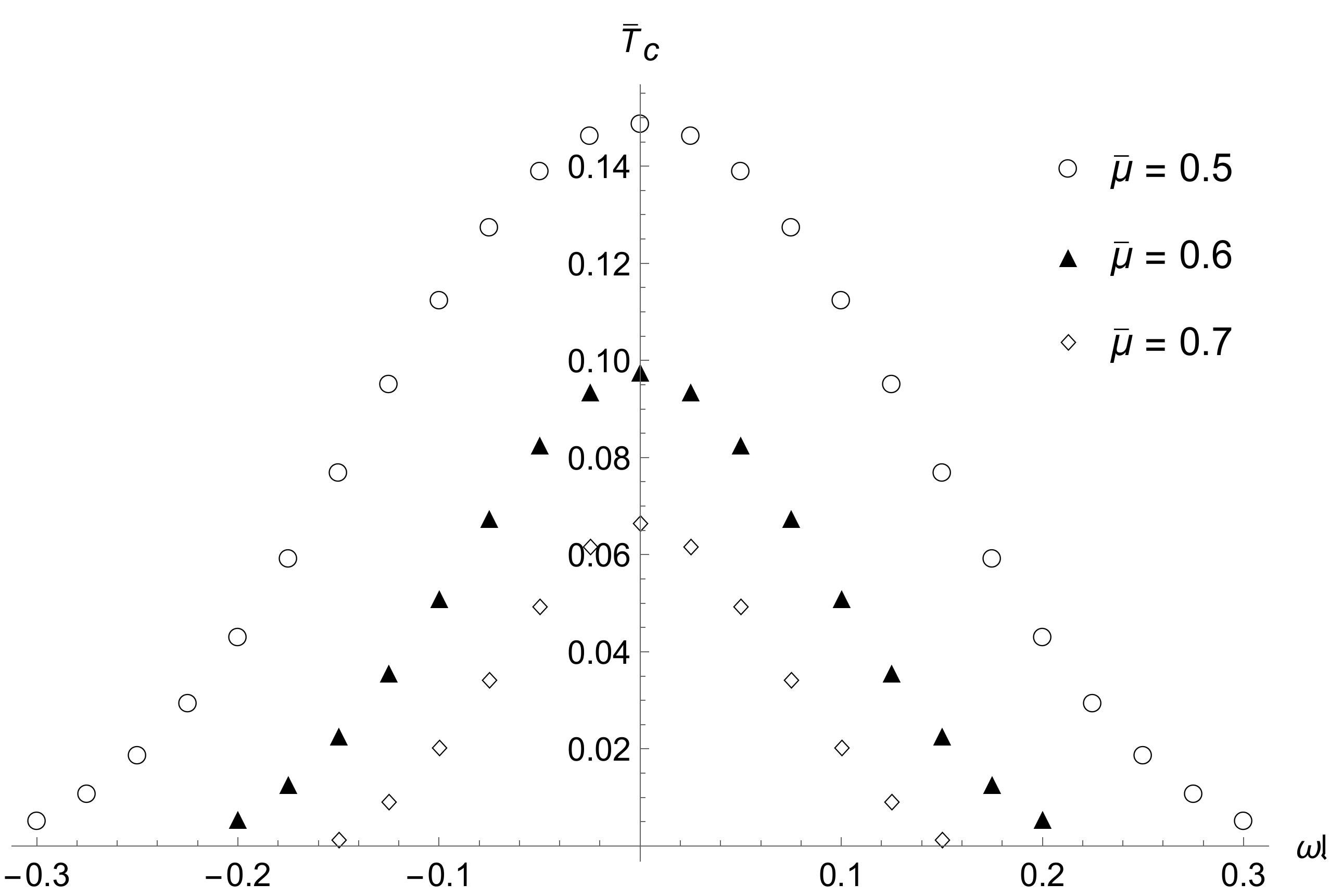}
	\caption{Critical temperature of deconfinement as a function of the plasma rotational velocity, at different values of quark chemical potential, in the exact Andree's model..}
    \label{fig3}
\end{figure}

\begin{figure}[!htb]
	\centering
	\includegraphics[scale=0.55]{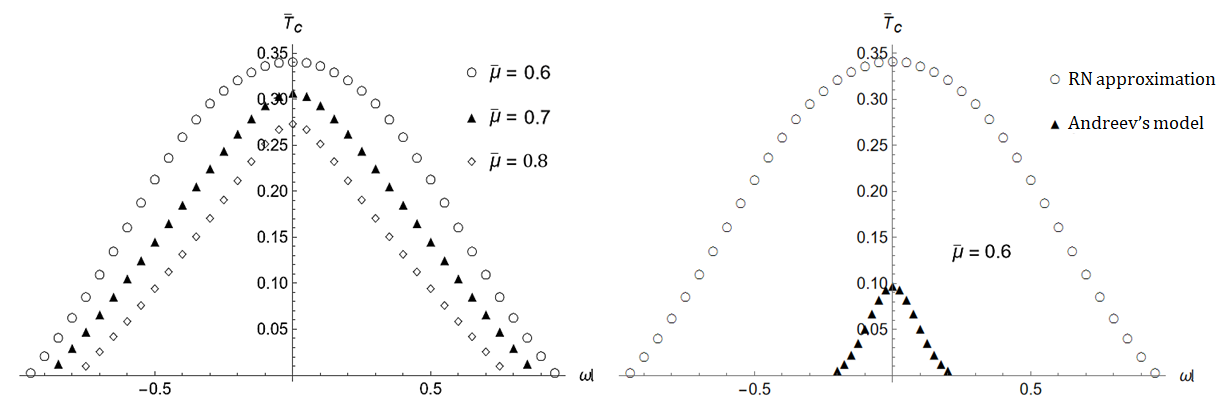}
	\caption{(Left) The same, in the RN approximation. (Right) Comparison between the exact and RN critical temperatures as a function of $\omega l$ at a fixed rotational velocity ($\omega l = 0.6)$.}
    \label{fig4}
\end{figure}
The values of $\omega_0$ depend on the quark chemical potential, $\omega_{0} = \omega_{0}(\bar{\mu})$, so that it decreases as $\bar{\mu}$ increases. This result imposes a strong constraint on the way hadronic matter can rotate at high values of the baryon density. In practice, $\omega_0$ corresponds to the value of the plasma angular velocity for which the phase transition occurs at $T_c = 0$. The numerical estimates of $\omega_0$ at different chemical potentials are displayed in Table \ref{table5} of the  Appendix, which was used to plot the curves in Fig. \ref{fig5}, for the exact and RN cases. As we can see, as the chemical potential (or the quark density) approaches zero, the rotational velocity $\omega_0 l \rightarrow 1$, and the constraint tends to disappear, with the hadronic matter being able to reach a rotation velocity comparable to the speed of light. The constraint is stronger in the exact case. For $\bar{\mu} \approx 0.15$ in the Andreev's model, the values of $\omega_0$ start to drop abruptly. On the other hand, when $\bar{\mu} \rightarrow 1.0675$, one observes that $\omega_0l \rightarrow 0$, which occurs when the quark chemical potential approaches the density corresponding to the most critical $\bar{\mu}$ of the phase transition at $T_c = 0$ in the case of a non-rotating QCD matter. In the RN approximation, it occurs when $\bar{\mu} \rightarrow 1.4142$.

\begin{figure}[!htb]
	\centering
	\includegraphics[scale=0.5]{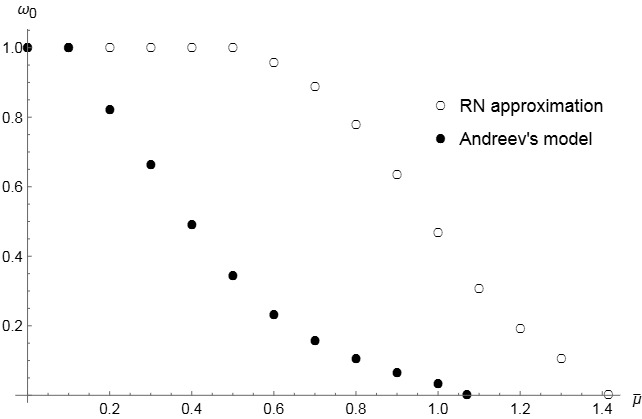}
	\caption{Most critical angular velocity of hadronic matter versus quark chemical potential, assuming $l=1$, in the exact Andreev's model and in the RN approximation.   The angular velocity is approaching the speed of light limit at small chemical potential.}
    \label{fig5}
\end{figure}

\section{QCD phase diagram,  most critical quark-chemical potential and its relation with rotation
}
\label{sec5}

\begin{figure}[!htb]
	\centering
	\includegraphics[scale=0.55]{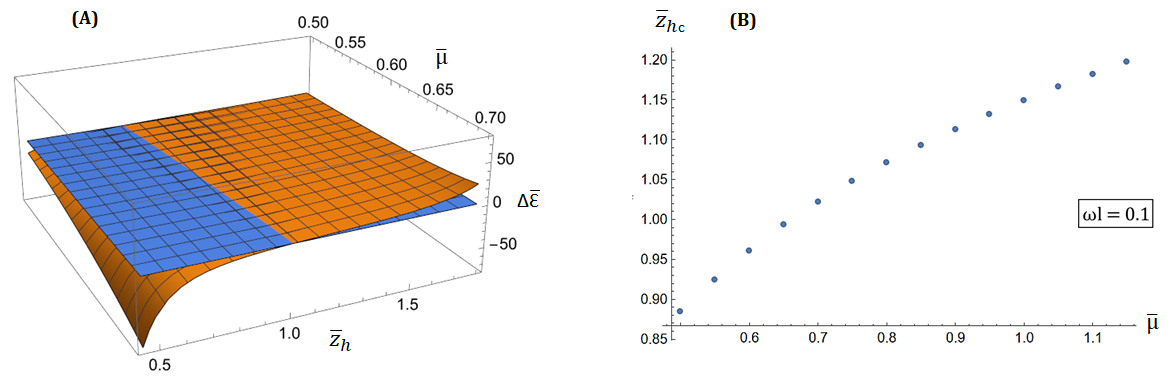}
	\caption{(A) Action density of regularized rotating BH as a function of $\bar{z}_h$ and $\bar{\mu}$, at a fixed rotational velocity $\omega l = 0.1$, with the intersection with the plane $\Delta \bar{\mathcal{E}} = 0$. (B) Critical horizon versus quark chemical potential at the same fixed $\omega l$.}
    \label{fig6}
\end{figure}

\begin{figure}[!htb]
	\centering
	\includegraphics[scale=0.54]{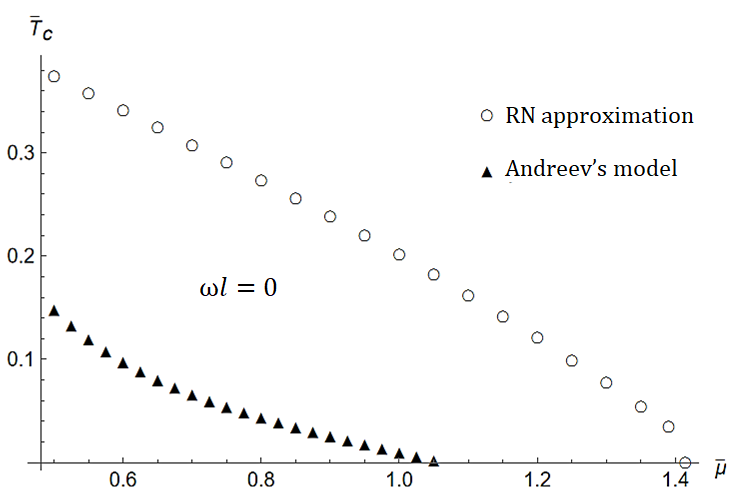}
	\caption{QCD phase diagram at high densities for a non-rotating matter: critical temperature versus quark chemical potential at $\omega l = 0$, in Andreev's exact soft wall model and in the RN approximation.}
    \label{fig7}
\end{figure}

\begin{figure}[!htb]
	\centering
	\includegraphics[scale=0.54]{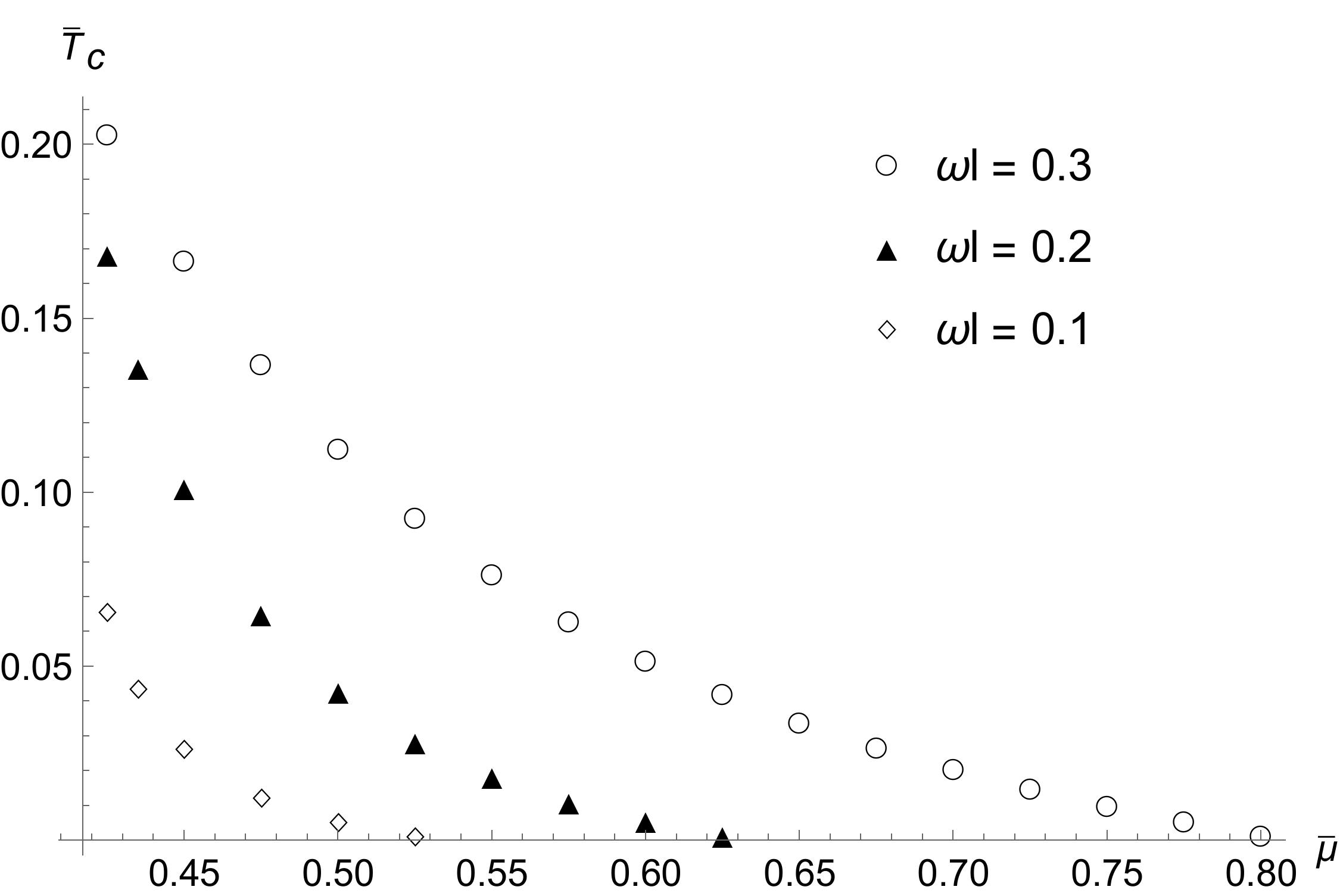}
	\caption{QCD phase diagram  for a rotating QCD matter: critical temperatures versus quark chemical potential at different rotational velocities, in the exact Andreev's model.}
    \label{fig8}
\end{figure}

The non-rotating QCD phase diagram at high densities in the exact and RN cases was plot in Fig. \ref{fig7}, according to the values of Table \ref{table6}. Applying the same procedure as in the previous section, we can use Eq. \eqref{HPtransition} to analyze the phase transitions at fixed non-zero rotational velocities, to obtain the critical temperatures as a function of $\bar{\mu}$ and $z_h$. In Fig. \ref{fig6}(A), we have plot $\bigtriangleup \bar{\mathcal{E}}$ as a function of $\omega$ and $z_h$ at $\omega l = 0.1$. The intersection between  $\bigtriangleup\bar{\mathcal{E}}(\bar{\mu},z_h)$ and $\bigtriangleup\bar{\mathcal{E}} = 0$ generates the curve of Fig. \ref{fig6}(B), whose  critical horizons were numerically computed, and can be found in Table \ref{table7}. For the  QGP analyzed at a given rotational velocity, the pattern remains always the same, see again Table \ref{table7}. From these numerical values, one generates the QCD phase diagrams at different rotational velocities in  Andreev's exact soft wall model, as displayed in Figs. \ref{fig7} and \ref{fig8}. 

The holographic prediction of Andreev's exact solution is in agreement with experimental data, where $T_c$ decreases as the quark chemical potential increases. The same behavior is observed for the rotating matter as illustrated by Fig. \ref{fig8}, where  $\bar{T}_c$ is plotted as a function of $\bar{\mu}$ at different rotational velocities. In both figures, one observes that there is a most critical quark chemical potential ($\mu_0$), above which phase transitions no longer occur. These chemical potentials correspond to the transition at $T_c = 0$. For $\mu \geq \mu_0$, the QCD matter is always in the deconfined plasma phase, independent of its temperature. The value of $\bar{\mu}_0$ depends on the angular velocity of the particles, $\bar{\mu}_0 = \bar{\mu}_0(\omega l)$, such that it decreases as $\omega l$ increases.    

\begin{figure}[!htb]
	\centering
	\includegraphics[scale=0.54]{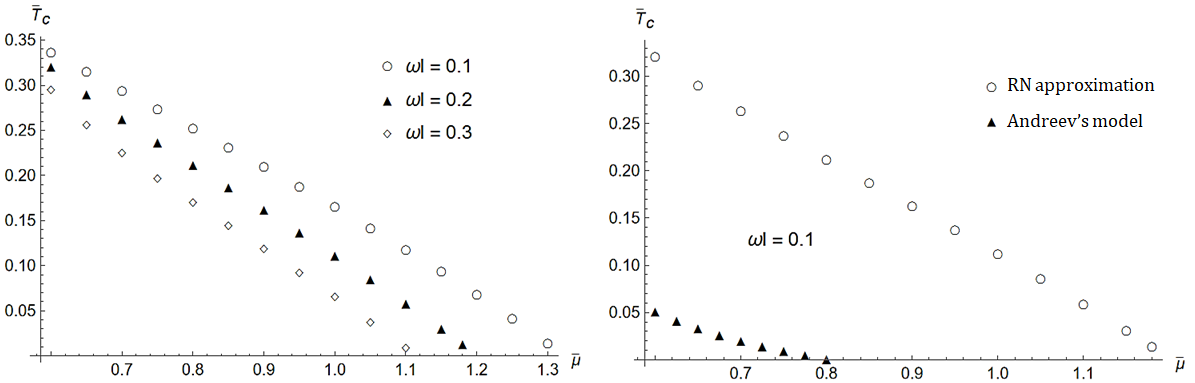}
	\caption{(Left) QCD phase diagram for a rotating QCD matter at different rotational velocities, in the RN approximation. (Right) Comparison between the critical temperatures of the exact and RN cases for a rotating matter with $\omega l = 0.1$. }
    \label{fig9}
\end{figure}

Qualitatively, the behavior of $T_c(\bar{\mu})$ and $\omega_0$ can be obtained in the RN approximation, see \cite{Braga:2025eiz}. However, the numerical values are not equivalent,  which demonstrates again that the contribution to the critical temperatures and densities from regions relatively distant from the bulk is not negligible, affecting the  phase diagram and causing a considerable reduction in the deconfinement temperatures. In Fig. \ref{fig9} (left), we have plot $\bar{T}_c(\bar{\mu})$ in the RN approximation, according to Tables \ref{table9} and \ref{table10}. These curves decrease more smoothly, but are given by much lower temperatures than in the exact case, see Fig. \ref{fig9} (right). In the soft-wall AdS/QCD model, bulk quantities such as the on-shell actions are integrated over the AdS radial coordinate $z$, which runs from the UV boundary ($z \to 0$) to the IR horizon ($z = z_h$). 
The RN approximation assumes that the near-boundary region ($z \ll 1/\sqrt{c}$) dominates, truncating the integrals where $e^{cz^2} \simeq 1 + cz^2$. 
In contrast, the exact Andreev solution includes exponential factors $e^{\pm cz^2}$ that enhance IR contributions. 
For large $z_h$ (low temperature or strong coupling), the deep-bulk region contributes significantly to the on-shell action: although $e^{-cz^2}$ suppresses the integrand, this is offset by the integration range and by factors of $z^n e^{cz^2}$. 
As a result, IR effects alter the quantitative values of the critical temperature and chemical potential, even if the qualitative behavior of $T_c(\mu,\omega)$ remains RN-like.

There are values of  $T_c$ and $\omega_0$ that were computed in regions with $z_{h c}$ not approximately one, where the RN approximation is not applicable, see Table \ref{table1} in the Appendix. For the non-rotating case, the numerical estimate of $\bar{\mu}_0$ at $T_c = 0$ in the exact model is  
\begin{equation}\label{barmu0}
    \bar{\mu}_0(0) = 1.0675\;. 
\end{equation}
For the rotational velocities analyzed in Fig. \ref{fig8}, one finds
\begin{eqnarray}
    \bar{\mu}_0(0.3) &=& 0.5342\;,\label{01234}\\
    \bar{\mu}_0(0.2) &=& 0.6374\;,\\
    \bar{\mu}_0(0.1) &=& 0.8069\;.\label{1234}
\end{eqnarray}
As we can see, the QCD phase diagram of Fig. \ref{fig8} and the curve of $\omega_0(\bar{\mu})$ in Fig. \ref{fig5} are complementary. As $\bar{\mu} \rightarrow \mu_0$, the most critical rotational velocity tends to zero, as
\begin{equation}
\lim_{\bar{\mu} \rightarrow 1.0675}\omega_0 l = 0\;, 
\end{equation}
as can be realized in Fig. \ref{fig5}, which corresponds to the most critical density for the transition to occur at $T_c = 0$, for a non-rotating plasma in the Andreev's model. Above this critical density, an infinitesimal increase in the angular momentum of the hadrons causes the matter to jump from the confined phase to the deconfined plasma one, irrespective of its temperature. 

\section{Final remarks and conclusions}
\label{sec6}

Using QCD phenomenology, one can fix the value of the IR parameter in the soft wall. The fit of the masses of lightest $\rho$-mesons leads to $\sqrt{c}=338\, \text{MeV}$ \cite{Herzog:2006ra}. In the exact Andreev's solution of the vector gauge field, there is a parameter $\eta$, which is considered to equal unity, for simplification. It affects the HP transitions in a non-trivial way \cite{Colangelo:2010pe}. To recover the dependence of the quark chemical potential on this parameter, one must define the physical density  
\begin{eqnarray}
    \bar{\mu}_{phys}(\eta) = \eta \bar{\mu}\;, 
\end{eqnarray}
where $\mu$ is the quark chemical potential with $\eta = 1$. For a QCD system with $N_c$ colors and $N_f$ flavors, the relation $Q = \eta q$ between the BH charge $Q$ and the metric parameter $q$  leads to the following expression of $\eta$ in holographic QCD models   \cite{Lee:2009bya}
\begin{equation}
    \eta  = \sqrt{\frac{3N_c}{2N_f}}\;.
\end{equation}

Taking into account a gauge group with $N_c = 3$ and $N_f = 6$ yields $\eta = \sqrt{3/4}$. This way, using Eq. \eqref{varsw}, considering the numerical estimate of $\bar{\mu}_0(0)$ \eqref{barmu0}, and the phenomenological value of $\sqrt{c}$, the physical Andreev's estimate for the most critical quark chemical potential is given by 
\begin{eqnarray}
    \mu_{0}(0) = \eta \sqrt{c}\bar{\mu}_0(0)  \approx \, 312.415 \, \text{MeV}\;. 
\end{eqnarray}
The relation between the quark and baryon chemical potentials is given by $\mu = \mu_B/3 $, see \cite{Colangelo:2010pe}, so that the holographic prediction of the most critical baryon density in the exact model is
\begin{eqnarray}\label{muB}
    \mu_{0B} (0)\approx 937.425 \, \text{MeV}\;. 
\end{eqnarray}
Comparing this result with the one obtained with the RN approximation\footnote{There is a slight difference in this estimate compared to \cite{Braga:2025eiz}. Here, we applied a more refined numerical method, which resulted in $\bar{\mu}^{RN}_0(0) \approx 1.4241$.}, see Ref. \cite{Braga:2025eiz},
\begin{eqnarray}
    \mu_{0 B}^{RN}(0) \approx 1250.570\,\text{MeV}\;,
\end{eqnarray}
one concludes that the contribution coming from regions relatively distant from the boundary leads to a significant reduction of the critical chemical potential. These predictions have the same order as the ones obtained in Einstein-Maxwell-dilaton (EMD) models  \cite{Chen:2020ath, Zhao:2022uxc}. The relation between the chemical potential for a rotating frame and the frame at rest is given by Eq. \eqref{muprime}. For a non-rotating QCD matter, the equality $ \mu^\prime_{0B} (0) =  \mu_{0B} (0)$ holds, which means that the result \eqref{muB} is general and does not depend on the observer.  

The rotation effects on the critical temperatures can be observed in Fig. \ref{fig3}, where we have plotted $T_c$ as a function of $\omega l$, at fixed quark chemical potentials. We have compared the exact and RN cases in Fig. \ref{fig4}, see Section \ref{sec4}. Our main result is displayed in Fig. \ref{fig5}, where we described  the behavior of the critical angular velocity $\omega_0 (\bar{\mu})$ at zero temperature in the exact Andreev's soft wall model and in the RN approximation, for a QCD matter rotating around a hypercylinder with unit radius. This curve corresponds to the points of Table \ref{table5} in the Appendix. To compute these points with precision, we need to access regions where the critical BH horizons are not approximately one, in which the RN approximation is not applied consistently. The difference of the $\omega_0(\bar{\mu})$ curves in the exact and RN cases is considerable, with the critical values of rotational velocities being lower in the exact model, see again Fig. \ref{fig5}. This reduction effect is interpreted as an influence from contributions coming from regions not near the AdS boundary, which cause a reduction in the deconfinement temperatures. They were computed using numerical methods applied to the HP phase transition equation \eqref{HPtransition} for a rotating charged BH, which does not have an analytical solution in the exact soft wall model.

In Section \ref{sec5}, we have plotted the QCD phase diagrams at high densities. Again, the description of $T_c(\bar{\mu})$ in the Hawking-Page approach for a rotating matter is in harmony with the one described from the computation of Polyakov loops in EMD models \cite{Chen:2020ath, Zhao:2022uxc}. As expected, at fixed rotational velocities, $\bar{T}_c$ decreases as $\bar{\mu}$ increases, see Figures \ref{fig7} and \ref{fig8}. In Fig. \ref{fig7}, we plot $\bar{T}_c(\bar{\mu})$ at $\omega l = 0$ in the exact and RN cases. The QCD phase diagram in the exact model for the rotating matter can be found in Fig. \ref{fig8}, and their results were compared with the RN approximation in Fig. \ref{fig9}.  As the density tends to its most critical value at zero temperature for the non-rotating matter, the most critical rotational velocity allowed for the hadronic matter tends to zero, see Fig. \ref{fig5}. In other words, for $\mu \geq \mu_0(0)$, any increase in the angular momentum of the hadrons would be enough to make them unstable, so that in this regime the QCD matter is always in the deconfined plasma phase. In this case, the hadrons could only be stable at absolute zero temperature, and with null angular momentum. From Fig. \ref{fig5}, we notice that $\omega_0$ starts to decrease abruptly at $\bar{\mu} \approx 0.15$, that is, for $\mu_{B} \approx 131.722 \, \text{MeV}$. In the most critical situation, $\omega_0 \rightarrow 0$ when $\bar{\mu} \rightarrow 1.0675$ (or $\mu_B \rightarrow 937.425\,\text{MeV}$).  Such predictions represent typical signatures of Andreev's exact soft wall model for a rotating QCD matter, which could be produced in noncentral heavy-ion collisions under critical conditions.          

\subsection*{Acknowledgements}
OCJ thanks The S\~ao Paulo Research Foundation (FAPESP) 
(Grants No. 2021/01089-1 and No. 2024/14390-0). The work of RdR is supported by The S\~ao Paulo Research Foundation (FAPESP) (Grants No. 2021/01089-1 and No. 2024/05676-7) and the National Council for Scientific and Technological Development - CNPq  (Grants No. 303742/2023-2 and No. 401567/2023-0).

\appendix

\section{Auxiliary tables with numerical results} \label{app1}

\begin{table}[h]
\centering
\begin{tabular}[c]{|c||c|c|c|}
\hline 
\diagbox{$\omega l$}{$\bar{\mu}$}  & $\,\,\,0.5\,\,\, $ & $\,\,\,0.6\,\,\, $ & $\,\,\,0.7\,\,\, $\\
 \hline\hline
 $\,\,\,0.000\,\,\, $ &$\,\,\,0.851777\,\,\,$ &$\,\,\,0.892832\,\,\,$&$\,\,\,0.922993\,\,\,$ \\
\hline
 $\,\,\,0.025\,\,\,$ & $\,\,\,0.853714\,\,\,$ &$\,\,\,0.897362\,\,\,$&$\,\,\,0.930231\,\,\,$ \\
\hline
 $\,\,\,0.050\,\,\,$ & $\,\,\,0.859589\,\,\,$ &$\,\,\,0.910687\,\,\,$&$\,\,\,0.950861\,\,\,$  \\
\hline 
 $\,\,\,0.075\,\,\,$ & $\,\,\,0.869578\,\,\,$  &$\,\,\,0.932087\,\,\,$&$\,\,\,0.982289\,\,\,$  \\
\hline
 $\,\,\,0.100\,\,\, $ & $\,\,\,0.883952\,\,\,$ &$\,\,\,0.960566\,\,\,$&$\,\,\,1.02162\,\,\,$  \\
\hline
 $\,\,\,0.125\,\,\,$ &$\,\,\,0.90303\,\,\,$ &$\,\,\,0.99503\,\,\,$&$\,\,\, 1.06637\,\,\,$  \\ 
\hline
 $\,\,\,0.150\,\,\,$ &$\,\,\,0.927117\,\,\,$ &$\,\,\,1.03439\,\,\,$&$\,\,\,1.11467\,\,\,$   \\
\hline 
$\,\,\,0.175\,\,\, $ &$\,\,\,0.956413\,\,\,$  &$\,\,\,1.07762\,\,\,$&$\,\,\,\textbf{x}\,\,\,$ \\
\hline
$\,\,\,0.200\,\,\,$ &$\,\,\, 0.990927\,\,\,$  &$\,\,\,1.12376\,\,\,$&$\,\,\,\textbf{x}\,\,\,$\\
\hline
 $\,\,\,0.225\,\,\,$ &$\,\,\,1.03040\,\,\,$  &$\,\,\,\textbf{x}\,\,\,$&$\,\,\,\textbf{x}\,\,\,$ \\ 
\hline
$\,\,\,0.250\,\,\,$ &$\,\,\,1.07428\,\,\,$  &$\,\,\,\textbf{x}\,\,\,$&$\,\,\,\textbf{x}\,\,\,$  \\
\hline  
$\,\,\,0.275\,\,\,$&$\,\,\,1.12178\,\,\,$  &$\,\,\,\textbf{x}\,\,\,$&$\,\,\,\textbf{x}\,\,\,$  \\
\hline
 $\,\,\,0.300\,\,\,$ &$\,\,\,1.17196\,\,\,$  &$\,\,\,\textbf{x}\,\,\,$&$\,\,\,\textbf{x}\,\,\,$ \\ 
\hline
 \end{tabular}   
\caption{Critical horizon positions in the Andreev's exact soft wall AdS/QCD model, at different chemical potentials ($\bar{\mu}$) and rotational velocities ($\omega l$), used to compute the temperatures of Table \ref{table2}. (The ``$\textbf{x}$" represents temperatures that do not obey the physical positivity condition of Eq. \eqref{positivity}, and were not used in Fig. \ref{fig3}.) }
\label{table1}
\end{table}

\begin{table}[h]
\centering
\begin{tabular}[c]{|c||c|c|c|}
\hline 
\diagbox{$\omega l$}{$\bar{\mu}$}  & $\,\,\,0.5\,\,\, $ & $\,\,\,0.6\,\,\, $ & $\,\,\,0.7\,\,\, $\\
 \hline\hline
 $\,\,\,0.000\,\,\, $ &$\,\,\,0.148783\,\,\,$ &$\,\,\,0.0980534\,\,\,$&$\,\,\,0.0665263\,\,\,$ \\
\hline
 $\,\,\,0.025\,\,\,$ & $\,\,\,0.146300\,\,\,$ &$\,\,\,0.0940484\,\,\,$&$\,\,\,0.0616726\,\,\,$ \\
\hline
 $\,\,\,0.050\,\,\,$ & $\,\,\,0.139023\,\,\,$ &$\,\,\,0.0831122\,\,\,$&$\,\,\,0.0493578\,\,\,$  \\
\hline 
 $\,\,\,0.075\,\,\,$ & $\,\,\,0.127461\,\,\,$  &$\,\,\,0.0679063\,\,\,$&$\,\,\,0.0342738\,\,\,$  \\
\hline
 $\,\,\,0.100\,\,\, $ & $\,\,\,0.112463\,\,\,$ &$\,\,\,0.0514437\,\,\,$&$\,\,\,0.0202849\,\,\,$  \\
\hline
 $\,\,\,0.125\,\,\,$ &$\,\,\,0.0951776\,\,\,$ &$\,\,\,0.0360873\,\,\,$&$\,\,\, 0.00922675\,\,\,$  \\ 
\hline
 $\,\,\,0.150\,\,\,$ &$\,\,\,0.0769580\,\,\,$ &$\,\,\,0.0231823\,\,\,$&$\,\,\,0.00139816\,\,\,$   \\
\hline 
$\,\,\,0.175\,\,\, $ &$\,\,\,0.0591928\,\,\,$  &$\,\,\,0.0131779\,\,\,$&$\,\,\,\textbf{x}\,\,\,$ \\
\hline
$\,\,\,0.200\,\,\,$ &$\,\,\, 0.0430901\,\,\,$  &$\,\,\,0.00593539\,\,\,$&$\,\,\,\textbf{x}\,\,\,$\\
\hline
 $\,\,\,0.225\,\,\,$ &$\,\,\,0.0294837\,\,\,$  &$\,\,\,\textbf{x}\,\,\,$&$\,\,\,\textbf{x}\,\,\,$ \\ 
\hline
$\,\,\,0.250\,\,\,$ &$\,\,\,0.0187373\,\,\,$  &$\,\,\,\textbf{x}\,\,\,$&$\,\,\,\textbf{x}\,\,\,$  \\
\hline  
$\,\,\,0.275\,\,\,$&$\,\,\,0.0107811\,\,\,$  &$\,\,\,\textbf{x}\,\,\,$&$\,\,\,\textbf{x}\,\,\,$  \\
\hline
 $\,\,\,0.300\,\,\,$ &$\,\,\,0.00524644\,\,\,$  &$\,\,\,\textbf{x}\,\,\,$&$\,\,\,\textbf{x}\,\,\,$ \\ 
\hline
 \end{tabular}    
\caption{Critical temperatures of deconfinement in the exact soft wall model at different chemical potentials and rotational velocities, corresponding to the points of Fig. \ref{fig3}. }
\label{table2}
\end{table} 

\begin{table}[h]
\centering
\begin{tabular}[c]{|c||c|c|c|}
\hline 
\diagbox{$\omega l$}{$\bar{\mu}$}  & $\,\,\,0.6\,\,\, $ & $\,\,\,0.7\,\,\, $ & $\,\,\,0.8\,\,\, $\\
 \hline\hline
 $\,\,\,0.00\,\,\, $ &$\,\,\,0.820006\,\,\,$ &$\,\,\,0.850888\,\,\,$&$\,\,\,0.877052\,\,\,$ \\
\hline
 $\,\,\,0.05\,\,\,$ & $\,\,\,0.821576\,\,\,$ &$\,\,\,0.857176\,\,\,$&$\,\,\,0.888259\,\,\,$ \\
\hline
 $\,\,\,0.10\,\,\,$ & $\,\,\,0.826351\,\,\,$ &$\,\,\,0.874736\,\,\,$&$\,\,\,0.917275\,\,\,$  \\
\hline 
 $\,\,\,0.15\,\,\,$ & $\,\,\,0.834534\,\,\,$  &$\,\,\,0.90080\,\,\,$&$\,\,\,0.956236\,\,\,$  \\
\hline
 $\,\,\,0.20\,\,\, $ & $\,\,\,0.846481\,\,\,$ &$\,\,\,0.932897\,\,\,$&$\,\,\,1.00009\,\,\,$  \\
\hline
 $\,\,\,0.25\,\,\,$ &$\,\,\,0.862725\,\,\,$ &$\,\,\,0.969503\,\,\,$&$\,\,\, 1.04659\,\,\,$  \\ 
\hline
 $\,\,\,0.30\,\,\,$ &$\,\,\,0.884008\,\,\,$ &$\,\,\,1.00992\,\,\,$&$\,\,\,1.09499\,\,\,$   \\
\hline 
$\,\,\,0.35\,\,\, $ &$\,\,\,0.911303\,\,\,$  &$\,\,\,1.05401\,\,\,$&$\,\,\,1.14526\,\,\,$ \\
\hline
$\,\,\,0.40\,\,\,$ &$\,\,\, 0.945817\,\,\,$  &$\,\,\,1.10197\,\,\,$&$\,\,\,1.19766\,\,\,$\\
\hline
 $\,\,\,0.45\,\,\,$ &$\,\,\,0.98892\,\,\,$  &$\,\,\,1.15426\,\,\,$&$\,\,\,1.25271\,\,\,$ \\ 
\hline
$\,\,\,0.50\,\,\,$ &$\,\,\,1.04199\,\,\,$  &$\,\,\,1.21148\,\,\,$&$\,\,\,1.31102\,\,\,$  \\
\hline  
$\,\,\,0.55\,\,\,$&$\,\,\,1.10618\,\,\,$  &$\,\,\,1.27443\,\,\,$&$\,\,\,1.37340\,\,\,$  \\
\hline
 $\,\,\,0.60\,\,\,$ &$\,\,\,1.18224\,\,\,$  &$\,\,\,1.34408\,\,\,$&$\,\,\,1.44086\,\,\,$ \\ 
\hline
 $\,\,\,0.65\,\,\,$ &$\,\,\,1.27059\,\,\,$  &$\,\,\,1.42168\,\,\,$&$\,\,\,1.51472\,\,\,$  \\
\hline 
 $\,\,\,0.70\,\,\,$ &$\,\,\,1.37174\,\,\,$ &$\,\,\,1.50898\,\,\,$&$\,\,\,1.59685\,\,\,$  \\
\hline
 $\,\,\,0.75\,\,\,$ &$\,\,\, 1.48703\,\,\,$ &$\,\,\,1.60858\,\,\,$&$\,\,\,1.69001\,\,\,$ \\
\hline
$\,\,\,0.80\,\,\,$ &$\,\,\,1.61979\,\,\,$   &$\,\,\,1.72483\,\,\,$&$\,\,\,\textbf{x}\,\,\,$ \\
\hline 
 $\,\,\,0.85\,\,\,$ &$\,\,\,1.77755\,\,\,$ &$\,\,\,1.86589\,\,\,$&$\,\,\,\textbf{x}\,\,\,$  \\
\hline
$\,\,\,0.90\,\,\,$ &$\,\,\,1.97833\,\,\,$ &$\,\,\,\textbf{x}\,\,\,$&$\,\,\,\textbf{x}\,\,\,$  \\
\hline
$\,\,\,0.95\,\,\,$ &$\,\,\,2.27921\,\,\,$ &$\,\,\,\textbf{x}\,\,\,$&$\,\,\,\textbf{x}\,\,\,$  \\
\hline
 \end{tabular}   
\caption{Critical horizon positions in the RN approximation, at different chemical potentials and rotational velocities, used to compute the temperatures of Table \ref{table4}.}
\label{table3}
\end{table} 

\begin{table}[h]
\centering
\begin{tabular}[c]{|c||c|c|c|}
\hline 
\diagbox{$\omega l$}{$\bar{\mu}$}   & $\,\,\,0.6\,\,\, $ & $\,\,\,0.7\,\,\, $ & $\,\,\,0.8\,\,\, $\\
 \hline\hline
 $\,\,\,0.00\,\,\, $ &$\,\,\,0.341197\,\,\,$ &$\,\,\,0.307734\,\,\,$&$\,\,\,0.273596\,\,\,$  \\
\hline
 $\,\,\,0.05\,\,\,$ & $\,\,\,0.339940\,\,\,$ &$\,\,\,0.304119\,\,\,$&$\,\,\,0.267540\,\,\,$ \\
\hline
 $\,\,\,0.10\,\,\,$ & $\,\,\,0.336160\,\,\,$ &$\,\,\,0.294194\,\,\,$&$\,\,\,0.252313\,\,\,$  \\
\hline 
 $\,\,\,0.15\,\,\,$ & $\,\,\,0.329833\,\,\,$  &$\,\,\,0.279911\,\,\,$&$\,\,\,0.232813\,\,\,$  \\
\hline
 $\,\,\,0.20\,\,\, $ & $\,\,\,0.320922\,\,\,$ &$\,\,\,0.263029\,\,\,$&$\,\,\,0.212042\,\,\,$  \\
\hline
 $\,\,\,0.25\,\,\,$ &$\,\,\,0.309382\,\,\,$ &$\,\,\,0.244690\,\,\,$&$\,\,\,0.191264\,\,\,$  \\ 
\hline
 $\,\,\,0.30\,\,\,$ &$\,\,\,0.295174\,\,\,$ &$\,\,\,0.225533\,\,\,$&$\,\,\,0.170908\,\,\,$   \\
\hline 
$\,\,\,0.35\,\,\, $ &$\,\,\,0.278287\,\,\,$  &$\,\,\,0.205898\,\,\,$&$\,\,\,0.151082\,\,\,$ \\
\hline
$\,\,\,0.40\,\,\,$ &$\,\,\, 0.258781\,\,\,$  &$\,\,\,0.185975\,\,\,$&$\,\,\,0.131779\,\,\,$\\
\hline
 $\,\,\,0.45\,\,\,$ &$\,\,\,0.236845\,\,\,$  &$\,\,\,0.165884\,\,\,$&$\,\,\,0.112966\,\,\,$ \\ 
\hline
$\,\,\,0.50\,\,\,$ &$\,\,\,0.212852\,\,\,$  &$\,\,\,0.145723\,\,\,$&$\,\,\,0.094619\,\,\,$  \\
\hline  
$\,\,\,0.55\,\,\,$&$\,\,\,0.187391\,\,\,$  &$\,\,\,0.125592\,\,\,$&$\,\,\,0.0767306\,\,\,$  \\
\hline
 $\,\,\,0.60\,\,\,$ &$\,\,\,0.161204\,\,\,$  &$\,\,\,0.105604\,\,\,$&$\,\,\,0.0593222\,\,\,$ \\ 
\hline
 $\,\,\,0.65\,\,\,$ &$\,\,\,0.135056\,\,\,$  &$\,\,\,0.0858924\,\,\,$&$\,\,\,0.0424466\,\,\,$  \\
\hline 
 $\,\,\,0.70\,\,\,$ &$\,\,\,0.0666048\,\,\,$ &$\,\,\,0.0224351\,\,\,$&$\,\,\,0.0261964\,\,\,$  \\
\hline
 $\,\,\,0.75\,\,\,$ &$\,\,\, 0.0852305\,\,\,$ &$\,\,\,0.0479122\,\,\,$&$\,\,\,0.0107191\,\,\,$ \\
\hline
$\,\,\,0.80\,\,\,$ &$\,\,\,0.0622233\,\,\,$   &$\,\,\,0.0300194\,\,\,$&$\,\,\,\textbf{x}\,\,\,$ \\
\hline 
 $\,\,\,0.85\,\,\,$ &$\,\,\,0.0406816\,\,\,$ &$\,\,\,0.0132118\,\,\,$&$\,\,\,\textbf{x}\,\,\,$  \\
\hline
$\,\,\,0.90\,\,\,$ &$\,\,\,0.0207259\,\,\,$ &$\,\,\,\textbf{x}\,\,\,$&$\,\,\,\textbf{x}\,\,\,$  \\
\hline
$\,\,\,0.95\,\,\,$ &$\,\,\,0.00283162\,\,\,$ &$\,\,\,\textbf{x}\,\,\,$&$\,\,\,\textbf{x}\,\,\,$  \\
\hline
 \end{tabular}   
\caption{Critical temperatures of deconfinement in the RN approximation at different chemical potentials and rotational velocities, corresponding to the points of Fig. \ref{fig4}. }
\label{table4}
\end{table} 

\begin{table}[h]
\centering
\begin{tabular}[c]{|c||c|c|}
\hline 
\text{}   $\,\,\,\bar{\mu}\,\,\, $ & $\,\,\,\omega_0 \, (\text{RN})\,\,\, $& $\,\,\,\omega_0 \,(\text{exact})\,\,\, $\\
 \hline\hline
 $\,\,\,0.0\,\,\, $ &$\,\,\,0.9999\,\,\,$ &$\,\,\,0.9999\,\,\,$ \\
\hline
  $\,\,\,0.1\,\,\, $ &$\,\,\,0.9999\,\,\,$&$\,\,\,0.9999\,\,\,$   \\
\hline
  $\,\,\,0.2\,\,\, $ &$\,\,\,0.9999\,\,\,$  &$\,\,\,0.8200\,\,\,$ \\
\hline
 $\,\,\,0.3\,\,\, $ &$\,\,\,0.9999\,\,\,$&$\,\,\,0.6633\,\,\,$  \\
\hline
  $\,\,\,0.4\,\,\, $ &$\,\,\,0.9999\,\,\,$ &$\,\,\,0.4911\,\,\,$ \\
\hline
 $\,\,\,0.5\,\,\, $ &$\,\,\,0.9999\,\,\,$&$\,\,\,0.3420\,\,\,$  \\
\hline
 $\,\,\,0.6\,\,\, $ &$\,\,\,0.9565\,\,\,$ &$\,\,\,0.2319\,\,\,$ \\
\hline
 $\,\,\,0.7\,\,\, $ &$\,\,\,0.8858\,\,\,$&$\,\,\,0.1559\,\,\,$  \\
\hline
 $\,\,\,0.8\,\,\, $ &$\,\,\,0.7765\,\,\,$&$\,\,\, 0.1030\,\,\,$  \\
\hline
 $\,\,\,0.9\,\,\, $ &$\,\,\,0.6345\,\,\,$&$\,\,\,0.0645\,\,\,$  \\
\hline
 $\,\,\,1.0\,\,\, $ &$\,\,\,0.4662\,\,\,$&$\,\,\, 0.0332\,\,\,$  \\
\hline
 $\,\,\,1.067\,\,\, $ &$\,\,\,-\,\,\,$&$\,\,\,0.0000\,\,\,$  \\
\hline
 $\,\,\,1.1\,\,\, $ &$\,\,\,0.3046\,\,\,$&$\,\,\,\textbf{x}\,\,\,$  \\
\hline
  $\,\,\,1.2\,\,\, $ &$\,\,\,0.1918\,\,\,$ &$\,\,\,\textbf{x}\,\,\,$ \\
\hline
 $\,\,\,1.3.\,\, $ &$\,\,\,0.1055\,\,\,$&$\,\,\,\textbf{x}\,\,\,$   \\
\hline
  $\,\,\,1.4142\,\,\, $ &$\,\,\,0.0000\,\,\,$&$\,\,\,\textbf{x}\,\,\,$   \\
\hline
 \end{tabular}   
\caption{Most critical angular velocities at different quark chemical potentials in the exact model and RN approximation, used to plot Fig. \ref{fig5}.}
\label{table5}
\end{table} 
\begin{table}[h]
\centering
\begin{tabular}[c]{|c|c|c||c|c|c|}
\hline 
$\bar{\mu}$\, (\text{RN})  & $\,\,\,\bar{z}_{h,c}\, \text{(RN)}\,\,\, $ & $\,\,\,\bar{T}_c\,\text{(RN)}\,\,\, $ & $\bar{\mu}\, (\text{exact}) $ &  $\,\,\,\bar{z}_{h,c}\,\text{(exact)}\,\,\, $& $\,\,\,\bar{T}_c\,\text{(exact)}\,\,\,$ \\
 \hline \hline
 $\,\,\,0.500\,\,\, $ &$\,\,\,0.784795\,\,\,$ &$\,\,\,0.374370\,\,\,$&$\,\,\,0.500\,\,\,$ &$\,\,\,0.851776\,\,\,$&$\,\,\,0.148783\,\,\,$ \\
\hline
 $\,\,\,0.550\,\,\,$ & $\,\,\, 0.802876\,\,\,$ &$\,\,\,0.357808\,\,\,$&$\,\,\,0.525\,\,\,$ &$\,\,\,0.863140\,\,\,$&$\,\,\,0.133448\,\,\,$  \\
\hline
 $\,\,\,0.600\,\,\,$ & $\,\,\,0.820006\,\,\,$ &$\,\,\,0.341197\,\,\,$&$\,\,\,0.550\,\,\,$ &$\,\,\,0.873767\,\,\,$&$\,\,\,0.120083\,\,\,$  \\
\hline 
 $\,\,\,0.650\,\,\,$ & $\,\,\,0.836035\,\,\,$  &$\,\,\,0.324520\,\,\,$&$\,\,\,0.575\,\,\,$&$\,\,\,0.883660\,\,\,$&$\,\,\,0.108376\,\,\,$  \\
\hline
 $\,\,\,0.700\,\,\, $ & $\,\,\,0.850888\,\,\,$ &$\,\,\,0.307734\,\,\,$&$\,\,\,0.600\,\,\,$ &$\,\,\,0.892831\,\,\,$ &$\,\,\,0.0980534\,\,\,$ \\
\hline
 $\,\,\,0.750\,\,\,$ &$\,\,\,0.864552\,\,\,$ &$\,\,\,0.290780\,\,\,$&$\,\,\,0.625\,\,\,$&$\,\,\,0.901312\,\,\,$&$\,\,\,0.0888861\,\,\,$   \\ 
\hline
 $\,\,\,0.800\,\,\,$ &$\,\,\,0.877052\,\,\,$ &$\,\,\,0.273596\,\,\,$&$\,\,\,0.650\,\,\,$ &$\,\,\,0.909137\,\,\,$&$\,\,\,0.0806795\,\,\,$   \\
\hline 
$\,\,\,0.850\,\,\, $ &$\,\,\,0.888446\,\,\,$  &$\,\,\,0.256115\,\,\,$&$\,\,\,0.675\,\,\,$ &$\,\,\,0.916350\,\,\,$ &$\,\,\,0.0732713\,\,\,$\\
\hline
$\,\,\,0.900\,\,\,$ &$\,\,\, 0.898805\,\,\,$  &$\,\,\,0.238278\,\,\,$&$\,\,\,0.700\,\,\,$&$\,\,\,0.922993\,\,\,$&$\,\,\,0.0665263\,\,\,$ \\
\hline
 $\,\,\,0.950\,\,\,$ &$\,\,\,0.908209\,\,\,$  &$\,\,\,0.220028\,\,\,$&$\,\,\,0.725\,\,\,$&$\,\,\,0.929111\,\,\,$ &$\,\,\,0.0603322\,\,\,$ \\ 
\hline
$\,\,\,1.000\,\,\,$ &$\,\,\,0.916741\,\,\,$  &$\,\,\,0.201315\,\,\,$&$\,\,\,0.750\,\,\,$ &$\,\,\,0.934745\,\,\,$&$\,\,\,0.0545959\,\,\,$  \\
\hline  $\,\,\,1.050\,\,\,$&$\,\,\,0.924481\,\,\,$  &$\,\,\,0.182095\,\,\,$&$\,\,\,0.7775\,\,\,$&$\,\,\,0.939939\,\,\,$  &$\,\,\,0.0492398\,\,\,$ \\
\hline
 $\,\,\,1.100\,\,\,$ &$\,\,\,0.931507\,\,\,$  &$\,\,\,0.162328\,\,\,$&$\,\,\,0.800\,\,\,$&$\,\,\,0.944728\,\,\,$ &$\,\,\,0.0441995\,\,\,$ \\ 
\hline
 $\,\,\,1.150\,\,\,$ &$\,\,\,0.937890\,\,\,$  &$\,\,\,0.141980\,\,\,$&$\,\,\,0.825\,\,\,$&$\,\,\,0.949151\,\,\,$&$\,\,\,0.0394211\,\,\,$  \\
\hline 
 $\,\,\,1.200\,\,\,$ &$\,\,\,0.943695\,\,\,$ &$\,\,\,0.121023\,\,\,$&$\,\,\,0.850\,\,\,$&$\,\,\,0.953237\,\,\,$ &$\,\,\,0.0348597\,\,\,$  \\
\hline
 $\,\,\,1.250\,\,\,$ &$\,\,\, 0.948982\,\,\,$ &$\,\,\,0.0994299\,\,\,$&$\,\,\,0.875\,\,\,$ &$\,\,\,0.957018\,\,\,$&$\,\,\,0.0304774\,\,\,$ \\
\hline
$\,\,\,1.300\,\,\,$ &$\,\,\,0.953804\,\,\,$   &$\,\,\,0.0771801\,\,\,$&$\,\,\,0.900\,\,\,$&$\,\,\,0.960520\,\,\,$&$\,\,\,0.0262422\,\,\,$  \\
\hline 
 $\,\,\,1.350\,\,\,$ &$\,\,\,0.958210\,\,\,$ &$\,\,\,0.0542541\,\,\,$&$\,\,\,0.925\,\,\,$&$\,\,\,0.963769\,\,\,$&$\,\,\,0.0221272\,\,\,$  \\
\hline
 $\,\,\,1.4142\,\,\,$ &$\,\,\,0.963323\,\,\,$ &$\,\,\,0.0237997\,\,\,$&$\,\,\,0.950\,\,\,$&$\,\,\,0.966786\,\,\,$  &$\,\,\,0.0181095\,\,\,$ \\
\hline
 $\,\,\,-\,\,\,$ &$\,\,\,-\,\,\,$ &$\,\,\,-\,\,\,$&$\,\,\,0.975\,\,\,$&$\,\,\,0.969593\,\,\,$  &$\,\,\,0.0141694\,\,\,$ \\
\hline
 $\,\,\,-\,\,\,$ &$\,\,\,-\,\,\,$ &$\,\,\,-\,\,\,$&$\,\,\,1.000\,\,\,$&$\,\,\,0.972206\,\,\,$  &$\,\,\,0.0102904\,\,\,$ \\
\hline
 $\,\,\,-\,\,\,$ &$\,\,\,-\,\,\,$ &$\,\,\,-\,\,\,$&$\,\,\,1.025\,\,\,$&$\,\,\,0.974644\,\,\,$  &$\,\,\,0.00645784\,\,\,$ \\
\hline
 $\,\,\,-\,\,\,$ &$\,\,\,-\,\,\,$ &$\,\,\,-\,\,\,$&$\,\,\,1.067\,\,\,$&$\,\,\,0.978381\,\,\,$  &$\,\,\,0.00009050\,\,\,$ \\
\hline
 \end{tabular}   
\caption{Critical temperatures in the exact model and RN approximation for the non-rotating matter, used to plot Fig. \ref{fig7}. }
\label{table6}
\end{table} 

\begin{table}[h]
\centering
\begin{tabular}[c]{|c||c|c|c|c|}
\hline 
\diagbox{$\bar{\mu}$}{$\omega l$}  & $\,\,\,0.1\,\,\, $ & $\,\,\,0.2\,\,\, $ & $\,\,\,0.3\,\,\, $\\
 \hline \hline
 $\,\,\,0.425\,\,\, $ &$\,\,\,0.817449\,\,\,$ &$\,\,\,0.837695\,\,\,$&$\,\,\,0.947759\,\,\,$  \\
\hline
 $\,\,\,0.435\,\,\, $ &$\,\,\,-\,\,\,$ &$\,\,\,0.863005\,\,\,$&$\,\,\,0.994843\,\,\,$  \\
\hline
 $\,\,\,0.450\,\,\,$ & $\,\,\, 0.840120\,\,\,$ &$\,\,\,0.897200\,\,\,$&$\,\,\,1.04867\,\,\,$  \\
\hline
 $\,\,\,0.475\,\,\,$ & $\,\,\,0.862383\,\,\,$ &$\,\,\,0.947251\,\,\,$&$\,\,\,1.11753\,\,\,$  \\
\hline 
 $\,\,\,0.500\,\,\,$ & $\,\,\,0.883952\,\,\,$  &$\,\,\,0.990927\,\,\,$&$\,\,\,1.17196\,\,\,$  \\
\hline
 $\,\,\,0.525\,\,\, $ & $\,\,\,0.904634\,\,\,$ &$\,\,\,1.02962\,\,\,$&$\,\,\,1.21738\,\,\,$   \\
\hline
 $\,\,\,0.550\,\,\,$ &$\,\,\,0.924319\,\,\,$ &$\,\,\,1.06422\,\,\,$&$\,\,\,\textbf{x}\,\,\,$  \\ 
\hline
 $\,\,\,0.575\,\,\,$ &$\,\,\,0.942961\,\,\,$ &$\,\,\,1.09541\,\,\,$&$\,\,\,\textbf{x}\,\,\,$    \\
\hline 
$\,\,\,0.600\,\,\, $ &$\,\,\,0.960566\,\,\,$  &$\,\,\,1.12376\,\,\,$&$\,\,\,\textbf{x}\,\,\,$ \\
\hline
$\,\,\,0.625\,\,\,$ &$\,\,\, 0.977171\,\,\,$  &$\,\,\,1.14972\,\,\,$&$\,\,\,\textbf{x}\,\,\,$ \\
\hline
 $\,\,\,0.650\,\,\,$ &$\,\,\,0.992835\,\,\,$  &$\,\,\,\textbf{x}\,\,\,$&$\,\,\,\textbf{x}\,\,\,$ \\ 
\hline
$\,\,\,0.675\,\,\,$ &$\,\,\,1.00763\,\,\,$  &$\,\,\,\textbf{x}\,\,\,$&$\,\,\,\textbf{x}\,\,\,$  \\
\hline  $\,\,\,0.700\,\,\,$&$\,\,\,1.02162\,\,\,$  &$\,\,\,\textbf{x}\,\,\,$&$\,\,\,\textbf{x}\,\,\,$ \\
\hline
 $\,\,\,0.725\,\,\,$ &$\,\,\,1.03488\,\,\,$  &$\,\,\,\textbf{x}\,\,\,$&$\,\,\,\textbf{x}\,\,\,$  \\ 
\hline
 $\,\,\,0.750\,\,\,$ &$\,\,\,1.04749\,\,\,$  &$\,\,\,\textbf{x}\,\,\,$&$\,\,\,\textbf{x}\,\,\,$  \\
\hline 
 $\,\,\,0.775\,\,\,$ &$\,\,\, 1.05950\,\,\,$ &$\,\,\,\textbf{x}\,\,\,$&$\,\,\,\textbf{x}\,\,\,$  \\
\hline
$\,\,\,0.800\,\,\,$ &$\,\,\,1.07098\,\,\,$   &$\,\,\,\textbf{x}\,\,\,$&$\,\,\,\textbf{x}\,\,\,$ \\
\hline 
 \end{tabular}   
\caption{Critical horizon positions in the exact soft wall model at different chemical potentials and rotational velocities, used to compute the temperatures of Table \ref{table7}.}
\label{table7}
\end{table} 

\begin{table}[h]
\centering
\begin{tabular}[c]{|c||c|c|c|c|}
\hline 
\diagbox{$\bar{\mu}$}{$\omega l$}  & $\,\,\,0.1\,\,\, $ & $\,\,\,0.2\,\,\, $ & $\,\,\,0.3\,\,\, $\\
 \hline \hline
 $\,\,\,0.425\,\,\, $ &$\,\,\,0.202753\,\,\,$ &$\,\,\,0.168666\,\,\,$&$\,\,\,0.0657364\,\,\,$  \\
\hline
$\,\,\,0.435\,\,\, $ &$\,\,\,-\,\,\,$ &$\,\,\,0.136177\,\,\,$&$\,\,\,0.0436076\,\,\,$  \\
\hline
 $\,\,\,0.450\,\,\,$ & $\,\,\, 0.166484\,\,\,$ &$\,\,\,0.101651\,\,\,$&$\,\,\,0.0263589\,\,\,$  \\
\hline
 $\,\,\,0.475\,\,\,$ & $\,\,\,0.136749\,\,\,$ &$\,\,\,0.065272\,\,\,$&$\,\,\,0.0123177\,\,\,$  \\
\hline 
 $\,\,\,0.500\,\,\,$ & $\,\,\,0.112463\,\,\,$  &$\,\,\,0.0430901\,\,\,$&$\,\,\,0.00524644\,\,\,$  \\
\hline
 $\,\,\,0.525\,\,\, $ & $\,\,\,0.0926051\,\,\,$ &$\,\,\,0.0285766\,\,\,$&$\,\,\,0.00110796\,\,\,$   \\
\hline
 $\,\,\,0.550\,\,\,$ &$\,\,\,0.0762839\,\,\,$ &$\,\,\,0.0185622\,\,\,$&$\,\,\,\textbf{x}\,\,\,$  \\ 
\hline
 $\,\,\,0.575\,\,\,$ &$\,\,\,0.0627613\,\,\,$ &$\,\,\,0.0113416\,\,\,$&$\,\,\,\textbf{x}\,\,\,$    \\
\hline 
$\,\,\,0.600\,\,\, $ &$\,\,\,0.0514437\,\,\,$  &$\,\,\,0.00593539\,\,\,$&$\,\,\,\textbf{x}\,\,\,$ \\
\hline
$\,\,\,0.625\,\,\,$ &$\,\,\, 0.0418631\,\,\,$  &$\,\,\,0.00175273\,\,\,$&$\,\,\,\textbf{x}\,\,\,$ \\
\hline
 $\,\,\,0.650\,\,\,$ &$\,\,\,0.0336545\,\,\,$  &$\,\,\,\textbf{x}\,\,\,$&$\,\,\,\textbf{x}\,\,\,$ \\ 
\hline
$\,\,\,0.675\,\,\,$ &$\,\,\,0.0265348\,\,\,$  &$\,\,\,\textbf{x}\,\,\,$&$\,\,\,\textbf{x}\,\,\,$  \\
\hline  $\,\,\,0.700\,\,\,$&$\,\,\,0.0202849\,\,\,$  &$\,\,\,\textbf{x}\,\,\,$&$\,\,\,\textbf{x}\,\,\,$ \\
\hline
 $\,\,\,0.725\,\,\,$ &$\,\,\,0.0147347\,\,\,$  &$\,\,\,\textbf{x}\,\,\,$&$\,\,\,\textbf{x}\,\,\,$  \\ 
\hline
 $\,\,\,0.750\,\,\,$ &$\,\,\,0.00975188\,\,\,$  &$\,\,\,\textbf{x}\,\,\,$&$\,\,\,\textbf{x}\,\,\,$  \\
\hline 
 $\,\,\,0.775\,\,\,$ &$\,\,\, 0.00523291\,\,\,$ &$\,\,\,\textbf{x}\,\,\,$&$\,\,\,\textbf{x}\,\,\,$  \\
\hline
$\,\,\,0.800\,\,\,$ &$\,\,\,0.00109641\,\,\,$   &$\,\,\,\textbf{x}\,\,\,$&$\,\,\,\textbf{x}\,\,\,$ \\
\hline 
 \end{tabular}      
\caption{Critical temperatures of deconfinement in the exact soft wall model at different chemical potentials and rotational velocities, corresponding to the points of Fig. \ref{fig8}. }
\label{table8}
\end{table}

\begin{table}[h]
\centering
\begin{tabular}[c]{|c||c|c|c|c|}
\hline 
\diagbox{$\bar{\mu}$}{$\omega l$}  & $\,\,\,0.1\,\,\, $ & $\,\,\,0.2\,\,\, $ & $\,\,\,0.3\,\,\, $\\
 \hline \hline
 $\,\,\,0.600\,\,\, $ &$\,\,\,0.826351\,\,\,$ &$\,\,\,0.846481\,\,\,$&$\,\,\,0.884008\,\,\,$  \\
\hline
 $\,\,\,0.650\,\,\, $ &$\,\,\,0.851241\,\,\,$ &$\,\,\,0.892811\,\,\,$&$\,\,\,0.955103\,\,\,$  \\
\hline
 $\,\,\,0.700\,\,\,$ & $\,\,\, 0.874736\,\,\,$ &$\,\,\,0.932897\,\,\,$&$\,\,\,1.00992\,\,\,$  \\
\hline
 $\,\,\,0.750\,\,\,$ & $\,\,\,0.896744\,\,\,$ &$\,\,\,0.968342\,\,\,$&$\,\,\,1.05559\,\,\,$  \\
\hline 
 $\,\,\,0.800\,\,\,$ & $\,\,\,0.917275\,\,\,$  &$\,\,\,1.00009\,\,\,$&$\,\,\,1.09499\,\,\,$  \\
\hline
 $\,\,\,0.850\,\,\, $ & $\,\,\,0.936395\,\,\,$ &$\,\,\,1.02877\,\,\,$&$\,\,\,1.12968\,\,\,$   \\
\hline
 $\,\,\,0.900\,\,\,$ &$\,\,\,0.954197\,\,\,$ &$\,\,\,1.05488\,\,\,$&$\,\,\,1.16067\,\,\,$  \\ 
\hline
 $\,\,\,0.950\,\,\,$ &$\,\,\,0.970792\,\,\,$ &$\,\,\,1.07879\,\,\,$&$\,\,\,1.18863\,\,\,$    \\
\hline 
$\,\,\,1.000\,\,\, $ &$\,\,\,0.986289\,\,\,$  &$\,\,\,1.10081\,\,\,$&$\,\,\,1.21410\,\,\,$ \\
\hline
$\,\,\,1.050\,\,\,$ &$\,\,\, 1.00080\,\,\,$  &$\,\,\,1.12120\,\,\,$&$\,\,\,1.23747\,\,\,$ \\
\hline
 $\,\,\,1.100\,\,\,$ &$\,\,\,1.01442\,\,\,$  &$\,\,\,1.14017\,\,\,$&$\,\,\,1.25904\,\,\,$ \\ 
\hline
$\,\,\,1.150\,\,\,$ &$\,\,\,1.02724\,\,\,$  &$\,\,\,1.15789\,\,\,$&$\,\,\,\textbf{x}\,\,\,$  \\
\hline  $\,\,\,1.200\,\,\,$&$\,\,\,1.03934\,\,\,$  &$\,\,\,\textbf{x}\,\,\,$&$\,\,\,\textbf{x}\,\,\,$ \\
\hline
 $\,\,\,1.250\,\,\,$ &$\,\,\,1.05081\,\,\,$  &$\,\,\,\textbf{x}\,\,\,$&$\,\,\,\textbf{x}\,\,\,$  \\ 
\hline
 $\,\,\,1.300\,\,\,$ &$\,\,\,1.06170\,\,\,$  &$\,\,\,\textbf{x}\,\,\,$&$\,\,\,\textbf{x}\,\,\,$  \\
\hline  
 \end{tabular}  
\caption{Critical horizon positions in the RN approximation at different chemical potentials and rotational velocities, used to compute the temperatures of Table \ref{table10}.}
\label{table9}
\end{table} 

\begin{table}[h]
\centering
\begin{tabular}[c]{|c||c|c|c|c|}
\hline 
\diagbox{$\bar{\mu}$}{$\omega l$}  & $\,\,\,0.1\,\,\, $ & $\,\,\,0.2\,\,\, $ & $\,\,\,0.3\,\,\, $\\
 \hline \hline
 $\,\,\,0.600\,\,\, $ &$\,\,\,0.336160\,\,\,$ &$\,\,\,0.320922\,\,\,$&$\,\,\,0.295174\,\,\,$  \\
\hline
 $\,\,\,0.650\,\,\, $ &$\,\,\,0.315109\,\,\,$ &$\,\,\,0.290500\,\,\,$&$\,\,\,0.256656\,\,\,$  \\
\hline
 $\,\,\,0.700\,\,\,$ & $\,\,\, 0.294194\,\,\,$ &$\,\,\,0.263029\,\,\,$&$\,\,\,0.225533\,\,\,$  \\
\hline
 $\,\,\,0.750\,\,\,$ & $\,\,\,0.273304\,\,\,$ &$\,\,\,0.237136\,\,\,$&$\,\,\,0.197507\,\,\,$  \\
\hline 
 $\,\,\,0.800\,\,\,$ & $\,\,\,0.252313\,\,\,$  &$\,\,\,0.212042\,\,\,$&$\,\,\,0.170908\,\,\,$  \\
\hline
 $\,\,\,0.850\,\,\, $ & $\,\,\,0.231092\,\,\,$ &$\,\,\,0.187249\,\,\,$&$\,\,\,0.144872\,\,\,$   \\
\hline
 $\,\,\,0.900\,\,\,$ &$\,\,\,0.209523\,\,\,$ &$\,\,\,0.162411\,\,\,$&$\,\,\,0.118880\,\,\,$  \\ 
\hline
 $\,\,\,0.950\,\,\,$ &$\,\,\,0.187500\,\,\,$ &$\,\,\,0.137277\,\,\,$&$\,\,\,0.0925926\,\,\,$    \\
\hline 
$\,\,\,1.000\,\,\, $ &$\,\,\,0.164931\,\,\,$  &$\,\,\,0.111657\,\,\,$&$\,\,\,0.0657719\,\,\,$ \\
\hline
$\,\,\,1.050\,\,\,$ &$\,\,\, 0.141734\,\,\,$  &$\,\,\,0.0854037\,\,\,$&$\,\,\,0.0382443\,\,\,$ \\
\hline
 $\,\,\,1.100\,\,\,$ &$\,\,\,0.117839\,\,\,$  &$\,\,\,0.0584020\,\,\,$&$\,\,\,0.00987816\,\,\,$ \\ 
\hline
$\,\,\,1.150\,\,\,$ &$\,\,\,0.0931846\,\,\,$  &$\,\,\,0.0305580\,\,\,$&$\,\,\,\textbf{x}\,\,\,$  \\
\hline  $\,\,\,1.200\,\,\,$&$\,\,\,0.0677189\,\,\,$  &$\,\,\,\textbf{x}\,\,\,$&$\,\,\,\textbf{x}\,\,\,$ \\
\hline
 $\,\,\,1.250\,\,\,$ &$\,\,\,0.0413953\,\,\,$  &$\,\,\,\textbf{x}\,\,\,$&$\,\,\,\textbf{x}\,\,\,$  \\ 
\hline
 $\,\,\,1.300\,\,\,$ &$\,\,\,0.0141735\,\,\,$  &$\,\,\,\textbf{x}\,\,\,$&$\,\,\,\textbf{x}\,\,\,$  \\
\hline  
 \end{tabular}     
\caption{Critical temperatures of deconfinement in the RN approximation at different chemical potentials and rotational velocities, corresponding to the points of Fig. \ref{fig9}. }
\label{table10}
\end{table}

\clearpage

\bibliographystyle{utphys2}
\bibliography{library}

\end{document}